\documentclass[10pt]{article}
\usepackage{graphicx}% Include figure files

\usepackage{epsfig}
\usepackage{latexsym}
\usepackage{amsmath}

\begin{document}

\title{%Socio-econo-physics of sects. \\
Benford's law and Theil transform\\
of % the Antoinist Cult
 financial data  }

\author{Paulette Clippe\footnote{
$e$-$mail$ $address$:
p.clippe@ulg.ac.be; {\it  previously at } GRAPES@SUPRATECS, Universit\'e
de Li\`ege,  B5 Sart-Tilman, B-4000 Li\`ege, Euroland}   
 \; and Marcel Ausloos\footnote{
$e$-$mail$ $address$:
marcel.ausloos@ulg.ac.be; {\it  previously at } GRAPES@SUPRATECS, Universit\'e
de Li\`ege,  B6 Sart-Tilman, B-4000 Li\`ege, Euroland
} \\   R\' esidence Beauvallon,
483 rue de la Belle Jardini\`ere, \\B-4031, Angleur, Belgium }

 \date{\today}
\maketitle \begin{abstract}
Among econophysics investigations,   studies of religious groups have been of interest.  On one hand, the present paper concerns the Antoinist community financial reports,  - a community which appeared at the end of the 19-th century in Belgium.   Several growth-decay regimes have been previously found over different time spans. However, there is common suspicion about   sect finances. In that spirit, the Antoinist community yearly financial reports, income and expenses,   are hereby examined along the so-called Benford's law. The latter is often used as a test about possible accounting wrongdoings. On the other hand,  Benford's law is known to be invariant under scale and base transformation. Therefore, as a further test,  of both such data and Benford's law use, the yearly financial reports are $ nonlinearly$ remapped through a sort of Theil transformation, i.e. based on a log-transformation. The resulting data is again  analyzed along the Benford's law scheme.   Bizarre, puzzling,  features are seen.  However, it is emphasized that such a non-linear transformation can shift the argument toward a more objective conclusion. In an Appendix, some brief discussion is made on why the original Theil mapping should not be used. In a second Appendix, an imperfect Benford's law-like form, - better suited for anomalous distributions, is presented.\end{abstract}

Keywords: income;
expenses;
religious community;
Benford's laws;
Theil map;
time series.

\maketitle

\section{Introduction  }\label{sec:intro}

The econophysics of religious movements and sects  has already attracted researchers 
\cite{PhA391.12.3190Ð97auslantoin,PicoliMendes08}, but  it could receive more attention. 
The finances, and more generally the economics,  of  religious movements  and  sects are often  questioned and the source of  major  political concerns, not discounting the ethics of  their portofolios and related matters \cite{DMW12302010}. Yet, Iannaccone  \cite{ianna95EI} has argued that many standard features of religious institutions exist to reduce (or at least appear to reduce) the risk of fraud and misinformation.
In pioneering work \cite{ianna98}, Iannaccone  has indicated the three  main lines
of research on such topics, and   has emphasizesd that  some religious behavior can be interpreted
from an economic perspective, applying microeconomic theory and techniques to explain  sect patterns among individuals, groups, and cultures.  This suggests an econophysics approach complementary to sociological ones.

The Antoinist community in Belgium and France  had much quick growth, in the number of adepts, but  not  so much anymore  \cite{Debouxhtaybook,Dericquebourg}.  It can be argued that money was not the cause of the growth \cite{Debouxhtaybook,Dericquebourg}, though the group got a legal   {\it Etablissement d'utilit\'e publique} (Organization of Public Utility) tax free status. In compensation, it had to publish
 financial reports in the official {\it Moniteur Belge} journal. Such a data for income and expenses is studied here below. The data acquisition is explained in Sect. \ref{sec:dataset}.

One technique to investigate the correct report of financial data is the use of Benford's law, outlined in Sect. \ref{sec:BL}.  Newcomb and later 
 Benford  \cite{ref[1],ref[2]} observed  that the occurrence of significant digits in many data sets is not
uniform but tends to follow a logarithmic distribution such that the smaller digits appear
as first significant digits more frequently than the larger ones, i.e.,  
\begin{equation}\label{Beneq1}
N_{d}= N\; log_{10}(1+\frac{1}{d}), \;\;\; d = 1, 2, 3, . . . , 9
\end{equation}
where $N$ is the total number of considered 1-st digits for  checking the law, in short, the number of data points,  and  $N_{d}$ is the number of the so observed integer $d$ ( $= 1, 2, 3, . . . , 9$)  being the starting one (1-st) in the data set list.

 Whence, the nowadays called Benford's law could  be used to identify  falsely created data, e.g. in corporationsÕ financial statements \cite{Nigrinibook}, or to
 verify  the (non)reliability of macroeconomic data of countries, - e.g., of too late interest in the recent case of Greece or Belgium \cite{GER12.11.243}. 

 In social sciences,  Benford's law has been
used to detect election frauds and anomalies, e.g. in USA or Iran \cite{ref[21],Roukema}. Closely related to our subject,
 Mir  investigated  whether regularities or anomalies exist in numerical data on the country-wise adherent distribution of seven major world religions  along Benford's law  \cite{TAMir}. 
 
%The latter has also been used in other cases in order to find out whether the data is reliable In physics the law has been successfully applied to numerical data on physical constants [10], atomic spectra [11], decay width of hadrons [12], magnitude and depth of earthquakes [6,13] and mantissa distribution of pulsars [14]. The law has been utilized in biological sciences to check the veracity of the data on clinical trials and discovery of drugs [15,16], study of diseases and genes [17]. The law is used in optimizing the size of computer files [18], enhancing the computing speed [19] and appearance of numbers on the internet [20].

When analyzing financial data, a transformation of the raw data can be made through what we call  a Theil map or Theil  transformation \cite{JM37,JM38,JM39,JM40}, 
in Sect. \ref{BenTheiltransf}. In order to compare  the   income $x_i$ of $M$ individuals  in a country over a time interval and thereby to  improve the  resolution for changes in high  incomes,
Theil used the index \cite{Theil65}
\begin{equation}
 \label{eq:IdA}
 Th = \frac{1}{M}\; \sum_i \frac{x_i}{\langle x_i \rangle} \ln \frac{x_i}{\langle x_i \rangle},
\end{equation}
summing $\frac{x_i}{\langle x_i \rangle}$ over the different times $i$ in the time interval and where  $\langle x_i \rangle $ is the  series mean value over the appropriate time interval.   However,  this induces negative and positive values of the (log-transformed) data, depending on the ratio $x_i/<x_i>$,  thus impairing a test of the validity  of Benford's law.
Therefore, in Sect. \ref{BenTheiltransf}, we will simply  transform the relevant data through the map\footnote{The Theil   index can be also transformed into a Theil entropy \cite{IdA}  
but this is not used here  either because it also induces negative and positive values of the (log-transformed) data,  thus impairing a test of the validity  of Benford's law.}  
\begin{equation}
 \label{eq:theil}
y_i\equiv\ x_i  \ln (x_i).
\end{equation}
Understand two arguments for doing so. On one hand, questions may be  raised whether Benford's law applies after a log-transform of the raw data \cite{Raimi76,PS8.11.1,MSHMSS186gauvrit09,CSM-349ScottFasli01},  
 while, on the other hand, one might also see some bizarre data reports through such a nonlinear transformation. 

Therefore, the paper goes as follows: as mentioned above, Sect. \ref{sec:BL} briefly  introduces  Benford's law.
In Sect. \ref{sec:dataset},  the raw data acquisition  is recalled, i.e. yearly  expenses and income over about 80 years, Sect.\ref{finAnt}. The data of interest is displayed (i) through histograms for  the first four digits, in Sect. \ref{Benrawdata}, and (ii) similarly  after a Theil transformation, in Sect. \ref{BenTheiltransf}. A discussion of such histograms is found in Sect. \ref{BLanal}.
 
Sect. \ref {sec:conclusions} serves for a conclusion emphasizing (i) the interest of  a Theil transformation to study data along Benford's law concepts, and (ii) the complexity of studying a community through its financial history.  

A brief discussion  of log-transformed data is given in  Appendix A. Moreover, since there is some apparent small deviation in the analyzed data, from the 1-st digit Benford law, an attempt to a better fit of the raw and Theil mapped data through a so called {\it imperfect 1-st- digit Benford law} is given in Appendix B.

\section{Benford's law}\label{sec:BL}

Benford's law \cite{ref[1],ref[2]},  Eq.(\ref{Beneq1}),  is  known as the  ÔÔfirst digit lawÕÕ or  the ÔÔlaw of the leading digitsÕÕ. According to Eq.(\ref{Beneq1}), in a given data set
the probability of occurrence of a certain digit as the first (1-st) significant digit decreases logarithmically as the value of the digit
increases from 1 to 9. Thus, digit 1 should appear as the first significant digit about 30.103\% times,  and
similarly 9 should appear about 4.576\% times.

 Benford's law, Eq.(\ref{Beneq1}),  holds for  data sets in which the occurrence of numbers is free from any restriction:  it does not apply to a list of telephone numbers, zip codes, ID-card numbers, car license plate numbers, .... It has been
found that tampered, unrelated or fabricated numbers usually do not follow Benford's law \cite{ref[44]}. Thus, significant deviations from
the Benford distribution may indicate fraudulent or corrupted data \cite{ref[45]}.   

On the other hand, much theoretical work has been done on the matter.  It has been discussed by many, in particular in \cite{Hill95,Hill96} that  base-invariance implies Benford's law. Whence, the law can be statistically derived along rigorous lines.

It is also possible to extend the law to digits beyond the first   \cite{Nigrini96}. In particular, the probability of encountering a number starting with a string of digits $n$ is given by 
\begin{equation}
    \log_{10}\left(n+1\right)-\log_{10}\left(n\right)=\log_{10}\left(1 +\frac{1}{n}\right)= \log_{10}\left( \frac{n+1}{n}\right),
\end{equation}
as its results from mere conditional probability algebra.\footnote {For example, the probability that a number starts with the digits 123 is $ log_{10}(1 + 1/123) = log_{10}(\frac{124}{123}) \;\sim 0.0035$.} This result can be used to find the probability that a particular digit occurs at a given position within a number.
E.g., the probability that $d$ ($d$ = 0, 1, ..., 9) is encountered as the $n$-th ($n$ $ >$  1) digit is
\begin{equation}
    \sum_{k=10^{n-2}}^{10^{n-1}-1} \log_{10}\left(1 + \frac{1}{10k+d}\right) =log_{10}\left[\;\prod_{k=10^{n-2}}^{10^{n-1}-1} (\frac{10k+d+1}{10k+d})\right].
\end{equation}
 For instance, the probability that a "2" is encountered as the second digit is  \cite{Nigrini96}  \begin{equation}   \log_{10}\left(1 +\frac{1}{12}\right)+\log_{10}\left(1 +\frac{1}{22}\right)+\cdots+\log_{10}\left(1 +\frac{1}{92}\right) \approx 0.1088. \end{equation}

The distribution of the $n$-th digit, as $n$ increases,  exponentially approaches a uniform distribution (UD)  
\cite{Nigrini96}. 
In practice, applications of Benford's law for fraud detection routinely use more than the first digit \cite{Nigrini96}. Indeed, the above can be generalized to forecast how many times any digit, or any combination, should be found at some rank in the number.
 The {\it ad hoc} table for the probable position of the digit $d$ up to rank 4 is given by Nigrini,
\cite{Nigrini96},  and requoted in \cite{Durtschietal04}. 

An extensive bibliography, from 1881 up to 2006,  on Benford's law papers including theories, applications, generalizations and warnings can be found in a H\"urlimann's  unpublished work \cite{0607168befordlawbiblio}.

\section{Antoinism community yearly budget data   }\label{sec:dataset}

\subsection{Financial data}\label{finAnt}
A community like the Antoinists exist for about more than a century  \cite{Debouxhtaybook,Dericquebourg}.  The time range of the financial data set  below  examined goes from 1922 till 2002,  i.e.   when it was mandatory to report the community finances. The data has been extracted from the  Belgian yearly official journal, the {\it Moniteur Belge}, when it was available in the archives of the Antoinist Cult Library in Jemeppe-sur-Meuse, Belgium. A few of these journals are missing, on various years but $ca.$ [1960-65] mainly,  without any known reason. However, this gap in data points does not appear {\it a posteriori}, i.e. after the subsequent data  analysis, to impair much the discussion and conclusion. Overall, there are 64 reports to be investigated.
 
 The budget data can be summarized in income and expenses. Note that this so called   "income"  value does not take into account the left-over from the pervious year(s).    A detailed study has led to interesting features about the finances evolutions \cite{PhA391.12.3190Ð97auslantoin}. 
 The raw data appears as  pretty scattered points. However, after much fit searching, it  has appeared that in both income and expenses cases, three growth-decay regimes can be found, with universal-like governing evolution laws.  Moreover the time interval limits are interpretable according to historical events \cite{PhA391.12.3190Ð97auslantoin}.   However, in [1], the time intervals in which a mathematically similar law is found for the  income or expenses data, are very slightly different from each other, as a result of fit optimization. As such, the best fits indicate mild variations in the border years of the three time interval regimes. 
 
For the present work,  it has been found convenient, from a mere statistical analysis point of view,   to choose an approximate border year value,  rather than the ones given in [1]. However, the same year, in both expenses and income cases, is used; see Table 1. 
In so doing,   the same number of data points exist  for expenses and income, whence allowing  to analyze the income and expenses data in the same  time intervals, 
 
 This "convenience argument" may be debated upon. We have the feeling, only the feeling, that, in view of the conclusions,   our argument should not lead to much controversy and can be scientifically accepted at this stage. 
 
 Note that  the data extends over  several orders of magnitudes, from about  7 10$^3$ BEF in 1922 till  above 7 $10^6$  BEF in 1976-77. Recall that the regimes present always a growth and a decay phase.
 
     \begin{table}\label{incomexpensregimes}\begin{center} \begin{tabular}{|c|c|c|c|c|c| l |    }
\hline     $expenses$&  $income$ &&$regime$ & & &  $N_I$ \\
\hline \hline $1922-1946 $  &   1922-1940 &&I&&1922-1940&14 \\
\hline $1946-1968$   &   1940-1968 &&II & &1941-1966&20\\
\hline $1968-1980  $  &1968-1980&&III&&1967-2001&30\\ 
\hline \end{tabular}  \end{center}
\caption{ Three growth phase regimes  found from best fits in \cite{PhA391.12.3190Ð97auslantoin}  on reported yearly expenses and  income data of the Belgium Antoinist community. The time intervals of the three regimes  found in \cite{PhA391.12.3190Ð97auslantoin} have been slightly adapted here, as explained in the main text,  for presently checking the Benford's law validity; $N_I$ is the corresponding number of data points  in the regime $I$  
}
\end{table}

 \subsection{Benford histogram of raw data}\label{Benrawdata}
 
 The Benford histogram for the 1-st   digit of  both  reported  income or expenses in the yearly budget of  the Belgian Antoinist community, the so called raw data is first examined. It is displayed in    Figs. 1(a)-2(a): % \ref{Fig1exp} - \ref{Fig2inc} 
     the previously found three growth-decay time interval regimes are distinguished, but also summed up in such  column stacking displays.  The  histograms for the 2-nd, 3-rd and 4-th digit are  shown in Figs. 1(b)-(d) and Figs. 2(b)-(d). % \ref{Fig1exp} - \ref{Fig2inc}. 
   The "theoretical", expected, Benford's law for the first two digits, Eq.(1) and Eq.(5) respectively,  are shown by darkened  triangles.  
   
The corresponding cases for the data resulting from  the  normalization, $x_i/<x_i>$, taking into account  the  average  $<x_i>$,  either over the full time interval or the respective $<x_i>$  in each regime (or  in each time interval), are not shown nor considered. Indeed, it is easily proved  \cite{Raimi76,PS8.11.1} that Benford's law is insensitive to such a change of scale, i.e., multiplying  (or dividing) the data by any positive scalar,  - here some average of the raw data over a time interval. Such an operation leads to identical   probability distribution functions of digits  \cite{Raimi76,PS8.11.1}.
 
 \subsection{Benford histogram of Theil transformed data}\label{BenTheiltransf}
 
 As discussed in the introduction, there are two reasons to investigate some transformed data along Benford's law concepts, i.e. (i) to predict theoretical differences,  when  a  non-linear transformation is used, and (ii) to observe whether one can deduce anomalies which would not be seen, if only  some raw data is analyzed.  Since it is known  \cite{Raimi76,PS8.11.1}  that a mere base or scale transformation is leaving Benford's law  invariant, it is of interest to examine more elaborate  transformations.

Recently,  theoretical  investigations on non-linear data transformation, like $log[log(x)]$, $\sqrt x$, $x^2$, have been reported \cite{MSHMSS186gauvrit09,CSM-349ScottFasli01}, though in another practical context. Our study along a Theil map,   Eq. (3), as done here, will thus serve as a complementary information to such publications.

The distributions of 1-st to 4-th digit of the transformed, expenses and income,  data according to $x\; log_{10} x$ are given in  the four subfigures of Figs. 3-4. A comment on the case $log[x/< x>]$ is found in Appendix A. 
 
 \subsection{Benford histograms analysis} \label{BLanal} 
 
 First, it  is  examined whether the distributions of the first digits  match the distribution specified by Benford's Law (BL), Eq.(1). Second, it is examined whether the first digits occur equally often at the second rank. 
 
 In order to do so,   $\chi^2$ tests have been used respectively on the 1-st and 2-nd digit through:
 
 \begin{equation}
 \label{eq:1BLchi2}
\chi^2_{1BL} =\sum_{i=1}^9 \frac{(d_{i1}-d_{i1B})^2}{d_{i1B}}
\end{equation}

 \begin{equation}
 \label{eq:2BLchi2}
\chi^2_{2BL}=\sum_{i=0}^9 \frac{(d_{i2}-d_{i2B})^2}{d_{i2B}},
\end{equation}
where $d_{i1B}$ and $d_{i2B}$ are the theoretically expected values according to BL. The results are given in Table 2.

For estimating the result interest, i.e. "feeling the numbers", one can make a comparable test of the data with respect to a simple uniform distribution (UD), i.e.

 \begin{equation}
 \label{eq:1UDchi2}
\chi^2_{1UD} =\sum_1^9 \frac{(d_{i1}-d_{1})^2}{d_{1}}
\end{equation}

 \begin{equation}
 \label{eq:2UDchi2}
\chi^2_{2UD}=\sum_0^9 \frac{(d_{i2}-d_{2})^2}{d_{2}}
\end{equation}
where $d_{1}$ and $d_{2}$ is the number expected from a uniform distribution at the 1-st and 2-nd rank respectively; here obviously, $d_{1}=64/9$ while $d_{2}= 64/10$. The results are given in Table 2.

First,  the fits to the Benford's law  for the  raw data and after the Theil mapping can be contrasted, for both cases, expenses or  income.  It appears that the BL $\chi^2$ value for the expenses test  is smaller in 7 cases out of 8;  the "not like others" case is the     1BL Theil map. This is exactly the opposite for the UD $\chi^2$ test.

  In the cases of 1BL and 1UD, these statistics may be compared to the $\chi^2$-distribution ($\chi^2_8$)
with 8 degrees of freedom. That distribution has a critical value of 15.5 for a 0.05-level test \cite{chi2table}. For 2BL and 2UD, these statistics may be compared to the $\chi^2$--distribution with 9 degrees of freedom ($\chi^2_9$), which
has a critical value of 16.9 for a 0.05-level test.  
 
 Because all of the statistics reported in Table 2  for  UD greatly exceed those values, the hypothesis that the   UD  is obeyed can be at once disregarded  \cite{CSM-349ScottFasli01}. The situation of BL is more surprising. Indeed, according to this goodness fit test, the 1BL does not seem to be obeyed for the raw data, niether for expenses nor income.  This is somewhat surprising. But one can observe that the 1-st digit is much more often 1 than "should be expected". In fact, the curvature of the envelope of the histogram even changes sign and curls up for the last digits, markedly smoothly in the case of the expenses, but in a less smooth way in the case of the income data. In both cases, the upturn occurs around digit 5. 
 Due to this observation, an attempt to generalize Benford's law taking into account such curling effect is given in Appendix B.
 
 However the BL holds for the 2-nd digit and for the Theil mapping, according to this goodness fit test. It is remarkable that already the distribution of the second digit  tends toward a flat distribution: the  "almost equally occurring digits" are 0,1,2,3.
 
We have not studied the distributions of the 3-rd and 4-th digit neither in the raw data nor in the Theil mapped data. Indeed, to find a $quasi$ uniform distribution for such cases  might not be expected due to the small number of data points \cite{ebeling280}. Fig. 3(d) is nevertheless remarkable, i.e. the 4-th digit distribution for the expenses raw data, indicates a sort of uniform background on which regular peaks are superposed. This bizarre behavior (of the 4-th digit !) might cast some doubt about the exactness of the content of the reported data   \cite{ref[44]}.

  Finally,  note that the Benford test  after the Theil mapping of the raw data gives $\chi^2$ values twice lower than for the raw data. In some sense one could have guessed such a feature for the 1BL. Indeed a mere logarithmic transformation would  induce an accumulation of first digits representing the power of 10. If the change in budget  is mild over several years, a peak would be found at some digit. However note that the transformation is $x\;log_{10}(x)$, thereby mixing digits. The $\chi^2$ test indicates some interesting though unknown up to now  feature of the Theil mapping for testing, e.g., faked data.

  \begin{table}\label{chi2Benford}\begin{center} \begin{tabular}{|c|c|c|c|c|c|c| l |    }
\hline $\chi^2$&&$expenses$   & $expenses$     & &$income$ & $income$  &   
 \\\hline
 \hline    &&1BL&   $2BL$&&1BL&$2BL$  &   \\
  \hline $raw$ $data   $ &&18.705  &  8.7351  &&   22.728&9.7651&  \\
\hline $Theil$ $ mapped$  &&  9.7386 & 4.9357   &&   7.1737 &5.3548  & \\ 
\hline\hline    &&1UD&   $2UD$&&1UD&$2UD$  &   \\
\hline   $raw$ $data$     &&35.528  &  49.529  &&  34.226&47.794&  \\
\hline $Theil$ $ mapped$  &&  46.999& 54.810   &&   50.963 &54.534 & \\ 
\hline \end{tabular}  \end{center}
\caption{ Values of the $\chi^2$ for a 1-st digit BenfordÕs Law (1BL) test  and 2nd-digit Benford's Law (2BL) test  of the raw and the Theil mapped data of reported yearly income and expenses of the Belgium Antoinist community  during the 20-th century.  The corresponding $\chi^2$ assuming a uniform distribution (UD) is given for comparison.   Recall the $\chi^2$-distribution critical value
  for a 0.05-level test \cite{chi2table}: $\chi^2_8 = 15.5 $
 and $\chi^2_9$= 16.9 with 8 and 9 degrees of freedom respectively }
%........  at the 0.01 level, the values are  20.09 and 21.67 respectively;
% at the 0.10 ... 13.36 and 14.68 
\end{table}
  
  % the use of a non-linear transformation in order
 
\section{Discussion}\label{sec:discussion}

As the starting point of a discussion, recall that the present work has two aims. One is a test of Benford's law on specific data:  the question being whether the data is reliable, or concurrently whether Benford's law is valid for such a case.  The second aim is  to touch upon the question whether Benford's law applies after a log-transform of the raw data, -  thus a non-linear transformation,  a question raised e.g., in  \cite{Raimi76,PS8.11.1,MSHMSS186gauvrit09,CSM-349ScottFasli01}. This is in line with recent considerations in statistical physics. Indeed, some time ago, physicists have been attracted  to study  financial time series and  to provide or present simple model equations  or algorithms for the data evolution; more exactly  making hypotheses on the $microscopic$ and $dynamical$ causes of the measured statistical characteristics;  e.g.  see \cite{stauffer04,chakr06,scienceculture}. 

The finances of  religious communities and sects in particular  
have been the object of rumors, scandals, criticisms, etc, for centuries and in various media or  groups. Some global data is sometimes freely available \cite{Evry,wikiLDSfin}, - however without any  (statiscally based) reliable control. Whence econo-physics tests and considerations are of interest.
Pay attention that finances of sects have been already discussed along statistical mechanics concepts \cite{PhA391.12.3190Ð97auslantoin,PicoliMendes08}. Ausloos looked at Antoinist cult community \cite{PhA391.12.3190Ð97auslantoin} because the data is legally available.   In \cite{PicoliMendes08}, partial distribution functions of adherents were found to follow laws as in economic activities and scientific research, suggesting that religious activities are governed by universal growth mechanisms.
However, one might  wonder whether the financial reports are faked.  

Thus, it has been argued, as here above, that it is of interest to analyze data along Benford's law lines. It is much agreed  that Benford's law  is  $definitively$ $not$  exact \cite{JudgeSchechter}. Thus, some  deviation from the Benford distribution would not provide a conclusive proof of manipulation, just as conformity does not prove cleanliness of the data.  Nevertheless, Benford's law(s) may  be considered useful as an aid in  analytical procedures of testing the exactness of  financial reports, like those of such religious movements  \cite{ref[45]}.    

Note that this empirical law has also been used in other cases in order to find out whether the data is reliable, as briefly mentioned by Mir \cite{TAMir} and by Pietronero \cite{PietroPhA293.01}. In physics and applied mathematics, the law has been applied to numerical data on physical constants \cite{[10]}, atomic spectra \cite{[11]},  values of radioactive decay half lives \cite{EJP14}, decay width of hadrons \cite{[12]}, magnitude and depth of earthquakes \cite{[13]}, mantissa distribution of pulsars \cite{[14]},  solutions or nonlinear differential equation systems   \cite{[119]}, or appearance of numbers on the internet \cite{[120]}.  The law  can be used in optimizing the size of computer files \cite{[18]} or enhancing the computing speed \cite{[19]}.

Thus, if the test is  conclusive, analyzers are happy, but if not, this induces more questions and reflexions. The apparent  lack of agreement with Benford's law for the 1-st digit of the raw data only, as found in the previous sections is somewhat frightening.  One could stop at this level, sending the Antoinists  treasurer and hierarchy to court for falsification. However an acceptable goodness fit test for the other cases, - disregarding the uniform distribution, of course, turns the case otherwise. This leads to basically two sets of questions,  economic and financial ones, about the specificity of the data, resulting from an accumulation of items : 
\begin{itemize} \item (i) Most likely, in order to resolve such a puzzle, one should demand more information on the items leading to the final sum of income and expenses. What was really  accounted for? Although the reported income and expenses pertained to a concluded year, it might occur that   some rounding factor accumulated so much in a few cases as modifying the first digits of items and finally the global report.  It does not mean that the Antoinist community has been cheating when reporting, - why should it? 
\item (ii) The anomalies might be only the result of sloppy, deliberate manipulation or unintentional but lazy accounting, quite in contrast from data manipulation by Governments \cite{GER12.11.243}.  
\item (iii) A third hypothesis might  have a more fundamental aspect: indeed, non-conformity with Benford's law should not be qualified as a reliable sign of poor quality of macroeconomic data, but could rather be based on marked structural shifts in the data set \cite{Gonzales-GarciaPastor09}. However, this is only throwing the stone toward closer inspection, much outside an econophysics investigation.
\end{itemize}

 In fact,  a subsequent question rests upon the measure of the tail of the distribution. It is known that it is not anomalous that the tail of a distribution function  often appears well measured on log-log plots.

At this stage of the conclusion, it should be emphasized that the Benford's law test  of  the resulting data after a Theil mapping of the raw data gives $\chi^2$ values  {\it in favor of } the Benford's law validity  for the Antoinists financial data reports. This $\chi^2$ test  result suggests to consider more often  the Theil mapping for testing whether there is some  faked data, within a Benford's law framework. This non-linear transformation should be further examined by mathematicians (and physicists) involved with research on the whereabouts of Benford's law.
 
 {\it In fine}, in light of theoretical and practical considerations along Theil's ideas, about income, \cite{Theil65} and recent work in macro- and micro-econophysics on the matter, it seems relevant to emphasize how useful it can be to observe data along various "lenses".

 \section{Conclusions}\label{sec:conclusions}
 
 A definite conclusion in scientific work is not always possible. We do stress that it would be regrettable that all conclusions of scientific papers be "definite conclusions".  Yet some practical information should be necessarily outlined after some  scientific analysis. The present approach has to be taken as leading to a warning on results, in some economic sense. The present financial case presents such an ambiguity, indeed, - at least at first sight. However, it seems that one method, as used here, i.e. a non-linear transformation of the data, can lead to more (or maybe less, in other cases) confidence on the data reliability.

In brief, two ingredients, yearly income and expenses,  of a religious movement financial reports have been here  investigated along  Benford's law concept. Moreover, we have also  transformed the reported data  through a non-linear mapping before  again testing Benford's law.  Note also that we have  been testing another distribution, the uniform distribution, beside the Benford one.  This allows  us  to have some convincing argument  about  the validity of the main approach.  The analysis of the    non-linearly transformed data strengthens the feeling toward an objective conclusion.
  
One practical conclusion, however, is that complementary accounting techniques  t to Benford's law should be considered before deciding whether some data is faked. Our attempt to use a non-linear transformation, through a sort of Theil map, suggests to consider a physicist technique. We doubt that it will be the case in accounting and political economy. But it might be used in scientific circles,  before being considered adequate in other fields.

\bigskip

{\bf Acknowledgements} Great thanks to the COST Action MP0801 for financial support.  Great thanks to reviewers for comments.  Correspondence with T.A. Mir is acknowledged by MA.

\bigskip

\section*{Appendix A: Log transform and Benford law}\label{sec:App_log}

One can directly take  the natural log of the yearly  income and expenses of the Belgium Antoinist community from the yearly budgets reported in the {\it Moniteur Belge}, i.e.  studying $log(x_i)$  digit distributions.
%However, as hinted in the main  text, after readily  examining the natural log of the raw, reported, yearly   income and expenses, a check of the Benford's law seems   unnecessary on the former data.

One can also "normalize"  the raw data, in order to define yearly relative income and expenses, $x_i/<x_i>$, taking into account  the average expenses and income, e.g. over the whole time range or within  the three time regimes, as done in Fig. \ref{fig:relincomexpenses}.   The overall relative  income and expenses data are also displayed. It is readily seen, as could be expected, that the corresponding normalization only results in a reduction of the older/low values of course. 

"Finally",  one could be  studying $log([x_i/<x_i>]$.   The transformed  data is shown in Fig.\ref{fig:lnrelatincomlnrelatexpenses}. 

Obviously, to divide by the average income or the average expenses merely result in a change of scale for the data. However, such a change  induces through the log transformation a set of values greater or smaller than zero (here, $ca.$ 1975)  which is incompatible with  the validity presumption of Benfor'd law. Nevertheless, note the accumulation of data points near integers, in particular  +2,  +1,  and also -1.

 For the same reasons, the  "official Theil mapping", Eq. (\ref{eq:IdA}), leading also to a set of negative terms,  was not  studied within this Benford's law framework. 
 
 Therefore, we have preferred studying the data resulting from the   non linear transformation $x_i\;log(x_i)$,   a sort of Theil mapping, Eq.(\ref{eq:theil}). 
 
\section*{Appendix B: Imperfect Benford's law}\label{sec:App_imperfectBL}

   It seems that raw data does not  always well agreed with Benford's law at $"large"$ 1st-digit ($d\ge5$). Sometimes a curl up feature can be seen, as in   Fig.1(a) and in Fig. 2(a). The exact reason is unknown. It might be due to psychological effects in business, i.e. it is better to pretend to have goods less expensive that they could be, thus reducing the sale price toward a lower decade. This increases the number of 9's in contrast to the number of 1's as a first digit. The same can be thought about tax evasion. Therefore, it is of interest to slightly modify  the 1st-digit Benford's law  to take such an effect into account.
   
   There are several ways of course to generalize Benford's law, keeping the writing as simple as possible, and with the same sort of analytical form. For completeness, let us point out that Benford's law was already "generalized" in a specific and {\it ad hoc} way \cite{FuShiSu} as
\begin{equation}\label{FuShiSu}
p_{r,q}(x) = N_{r,q}\; log_{10} (1+ \frac{1}{r+x^q}),
\; x= 1, 2,...,9 
\end{equation}
where $N_{r,q}$ is a normalization factor which makes $p_{r,q}(x)$ a probability distribution; $r$ and $q$ are model parameters to precisely
describe the distribution for different cases.   Obviously, when $r=0$ and $q= 1$,
one recovers the Benford's law distribution. Note that if  $r=1$, one might reanalyze other data sets, like zip code lists, telephone number lists, ...  for which the first digit can be 0. Thus, in Eq.(\ref{FuShiSu}), one would let $x \in  [0,9]$ without any log-divergence problem. 

However, the distribution curvature is unchanged when using Eq.(\ref{FuShiSu}). In order to have a curl up, i.e. a positive slope at "high" 1-st digit, another approach is mandatory. In order to keep the same analytic log-form,  as Eq.(1), we have found out that the most simple form is 
\begin{equation}\label{PCMA}
p_s(x) = N_s\; log_{10} ( \frac{1}{x}+ 1 + s\;Éx),
\; x= 1, 2,...,9 
\end{equation}
where $N_s$ is a normalization factor which makes $p_s(x)$ a probability distribution; $s$ being a model parameter to precisely fit the data. The minimum  of the distribution, $log_{10}(1+2\sqrt s)$, occurs at $1/\sqrt s $. Obviously, when $s=0$, one recovers the  Benford's law distribution. For illustration, see 
Fig.\ref{fig:Plot7colBJIimperfBen1}.

As a test, the 1st-digit histogram of the raw and Theil mapped data for both income and expenses of the reported Antoinist community budget are shown in Fig. 8.
%\ref{fig:FigAppB}. 
For ease in the fit, we have let $N_s$ be constrained to be an integer. In so doing the total surface $S$ of the histogram is not exactly equal to the number of data points, i.e. 64. The $s$, $N_s$,  $S$, and $\chi^2$ values  for the best fits are given in Table 3. Those values of course depend on the accuracy with which one constrains the nonlinear fits. Nevertheless, the fits are well improved in the case of the expenses. Again,  the Theil mapping markedly improves the overall fits.

%\newpage

  \begin{table}\label{ImperfBenford}\begin{center} \begin{tabular}{|c|c|c|c|c|c|c| l |    }
\hline  &&$expenses$   & $expenses$     & &$income$ & $income$  &   
 \\  \hline    &&$raw $&   $Theil$&&$raw$&$Theil$  &   \\
  \hline $N_s $ &&61 &  63  &&   62&63&  \\
\hline $s$  &&  0.0031 & 0.0012  &&   0.0021 &0.0011& \\ 
\hline   $\chi^2$     &&16.11&  8.58  &&  22.73&7.08&  \\
\hline $S$  &&  64.09& 64.24   &&64.13 &64.14 & \\ 
\hline \end{tabular}  \end{center}
\caption{ Values of the "Imperfect Benford's Law" parameters, $N_s$ and $ s$,  obtained for the fits with Eq.(\ref{PCMA}), to the raw or Theil mapped reported yearly income and expenses of the Belgium Antoinist community  during the 20-th century.  The  $\chi^2$  result is given together with the resulting value of the theoretical histogram surface $S$ } 
\end{table}

\newpage

 \begin{figure}
\centering
\includegraphics[height=6.0cm,width=6.0cm]{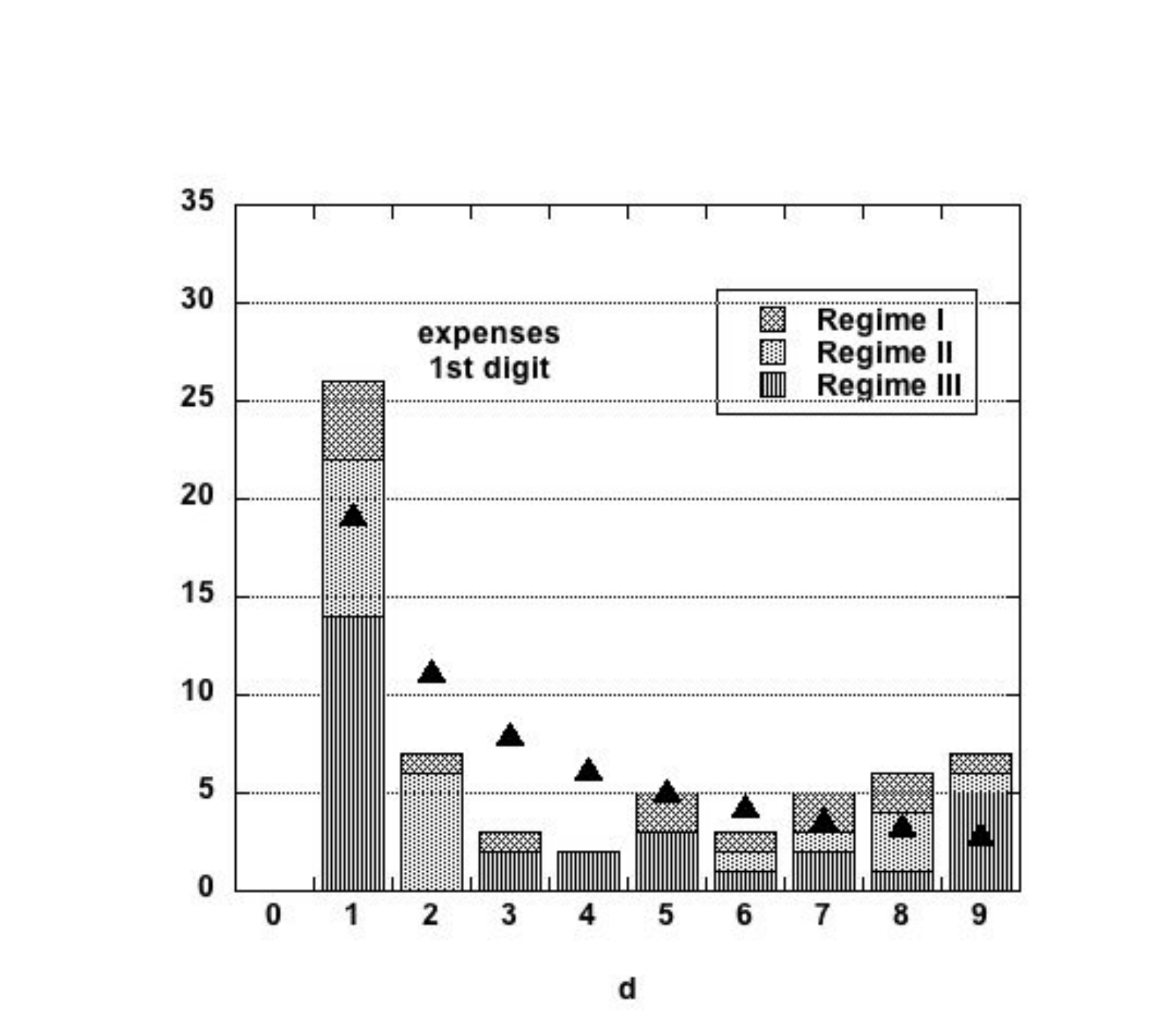}
\includegraphics[height=6.0cm,width=6.0cm]{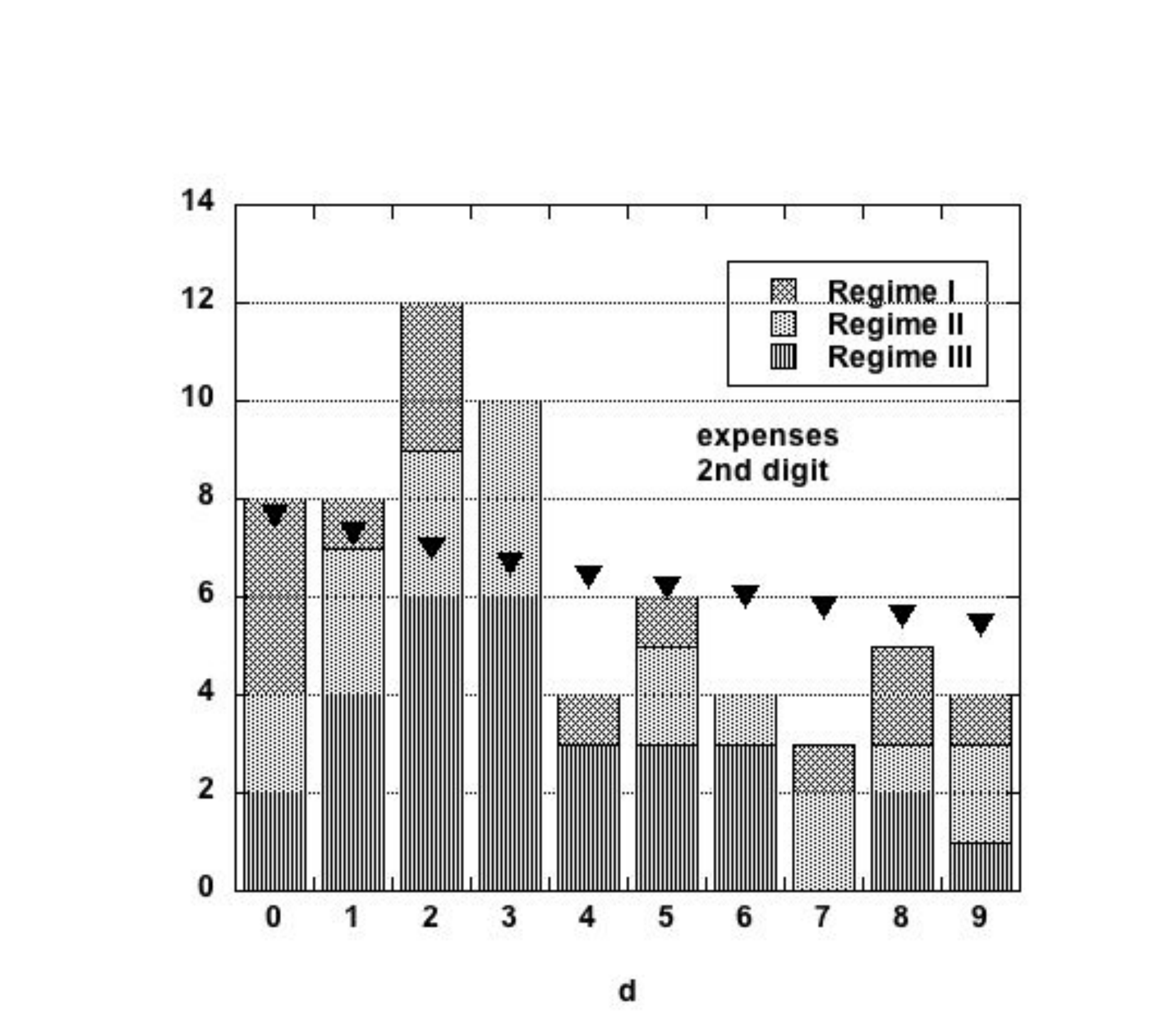}
\includegraphics[height=6.0cm,width=6.0cm]{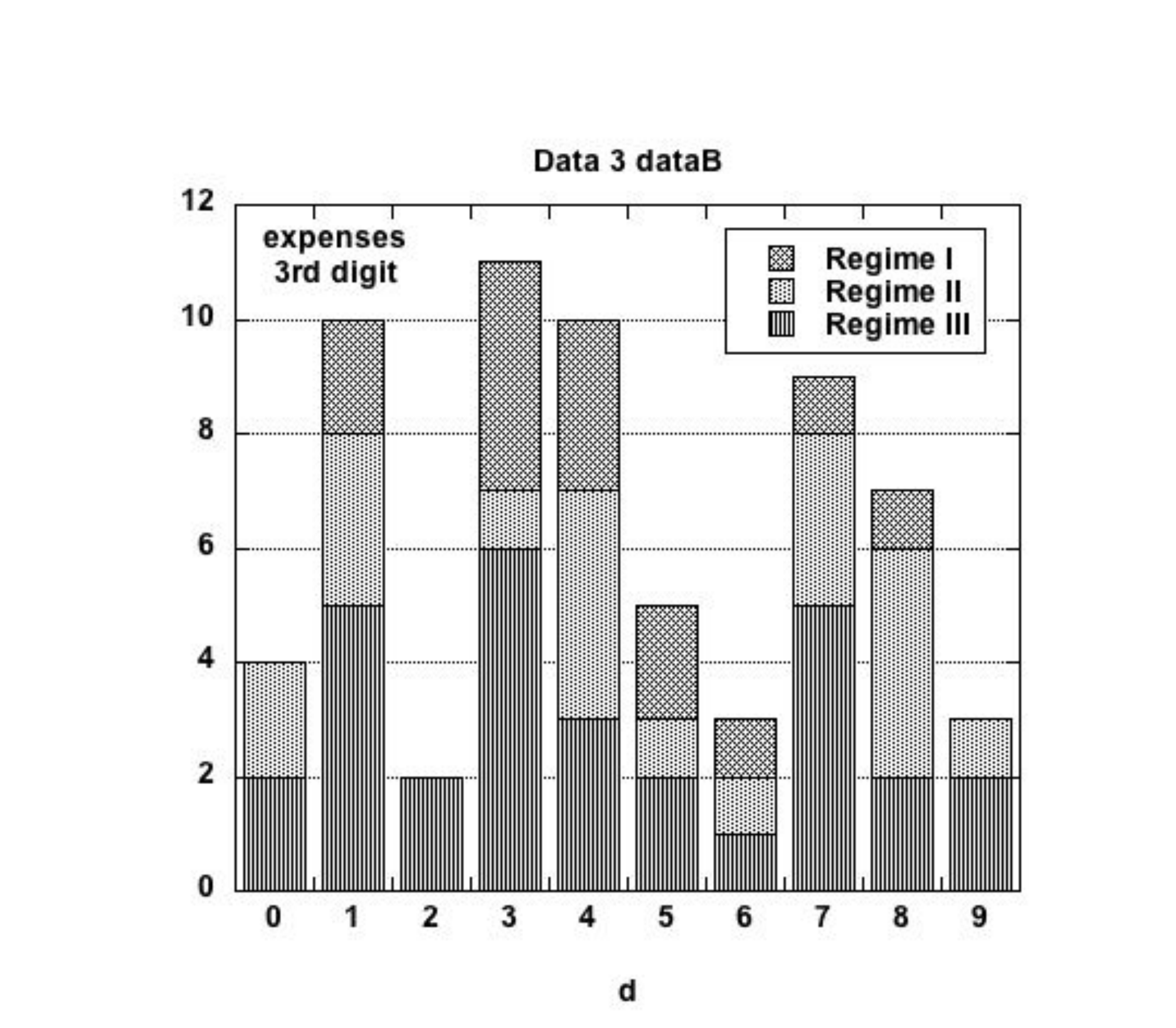}
\includegraphics[height=6.0cm,width=6.0cm]{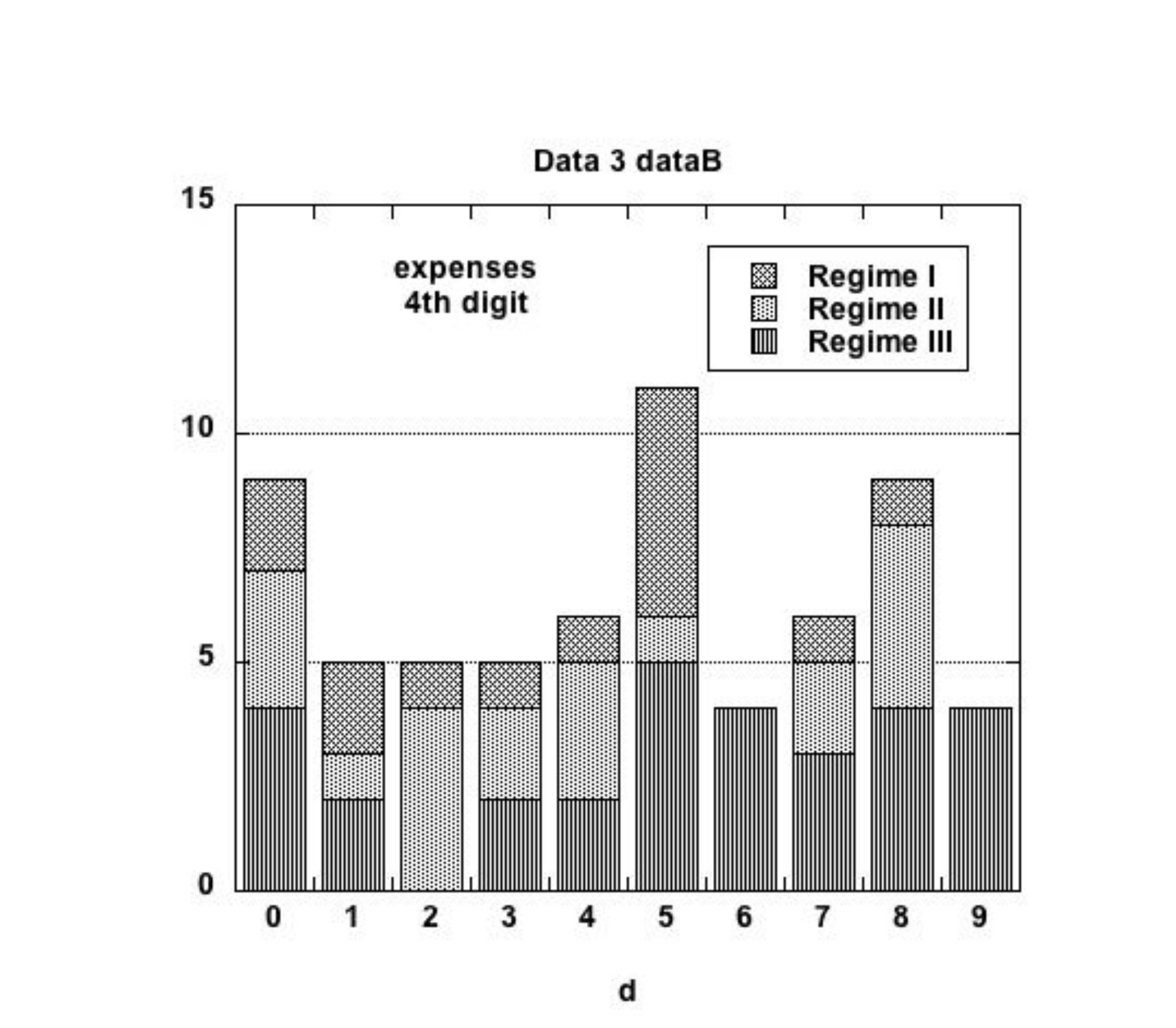}
\caption   {Benford histogram of the first four digits  of reported  expenses in the yearly budget of  the Belgian Antoinist community: the previously found three growth-decay time interval regimes are distinguished and summed up.  The expected Benford's laws for the first two digits are shown by  triangles} 
 \label{fig:Fig1exp} 
\end{figure}

 \begin{figure} 
\centering
\includegraphics[height=6.0cm,width=6.0cm]{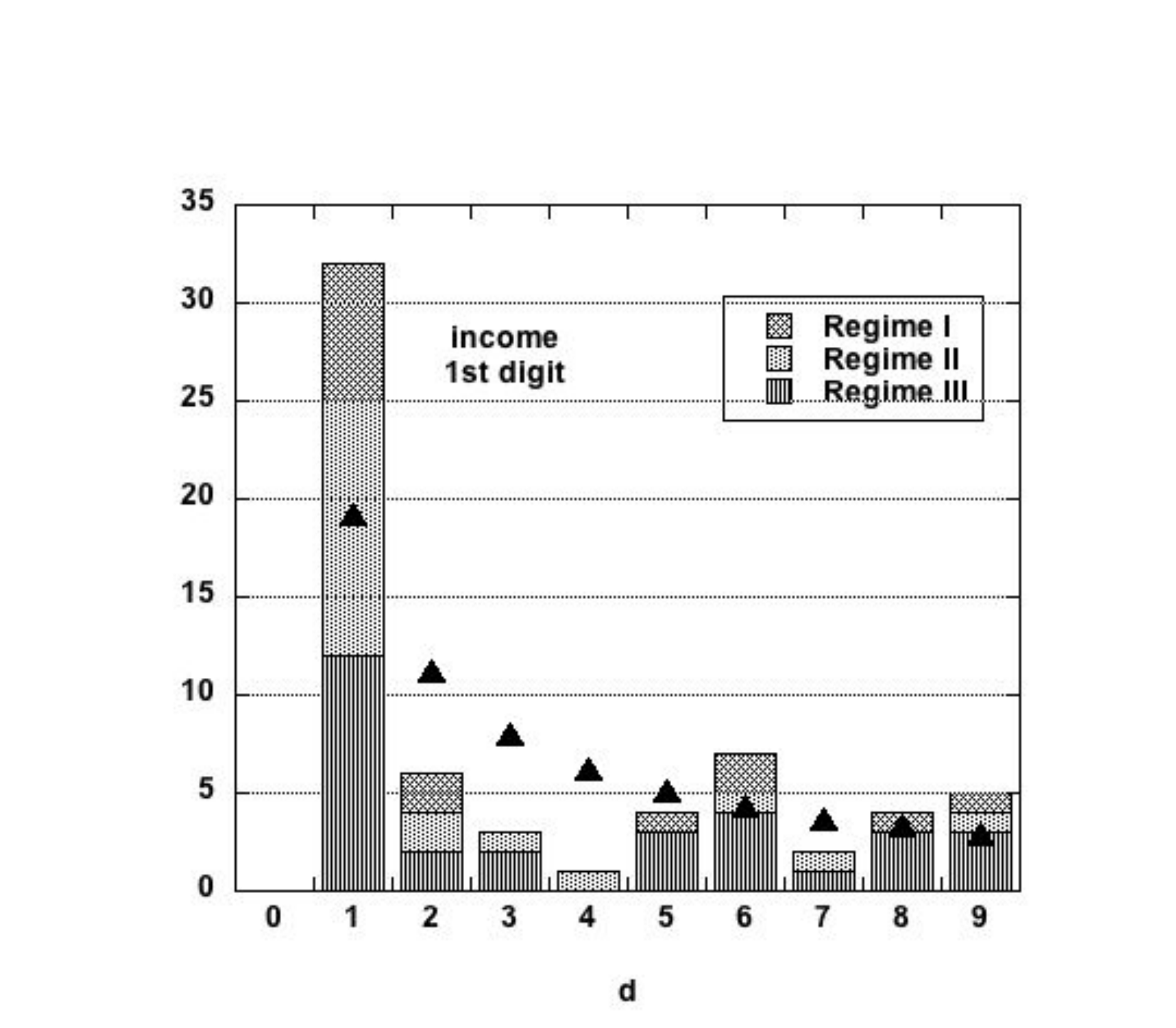}
\includegraphics[height=6.0cm,width=6.0cm]{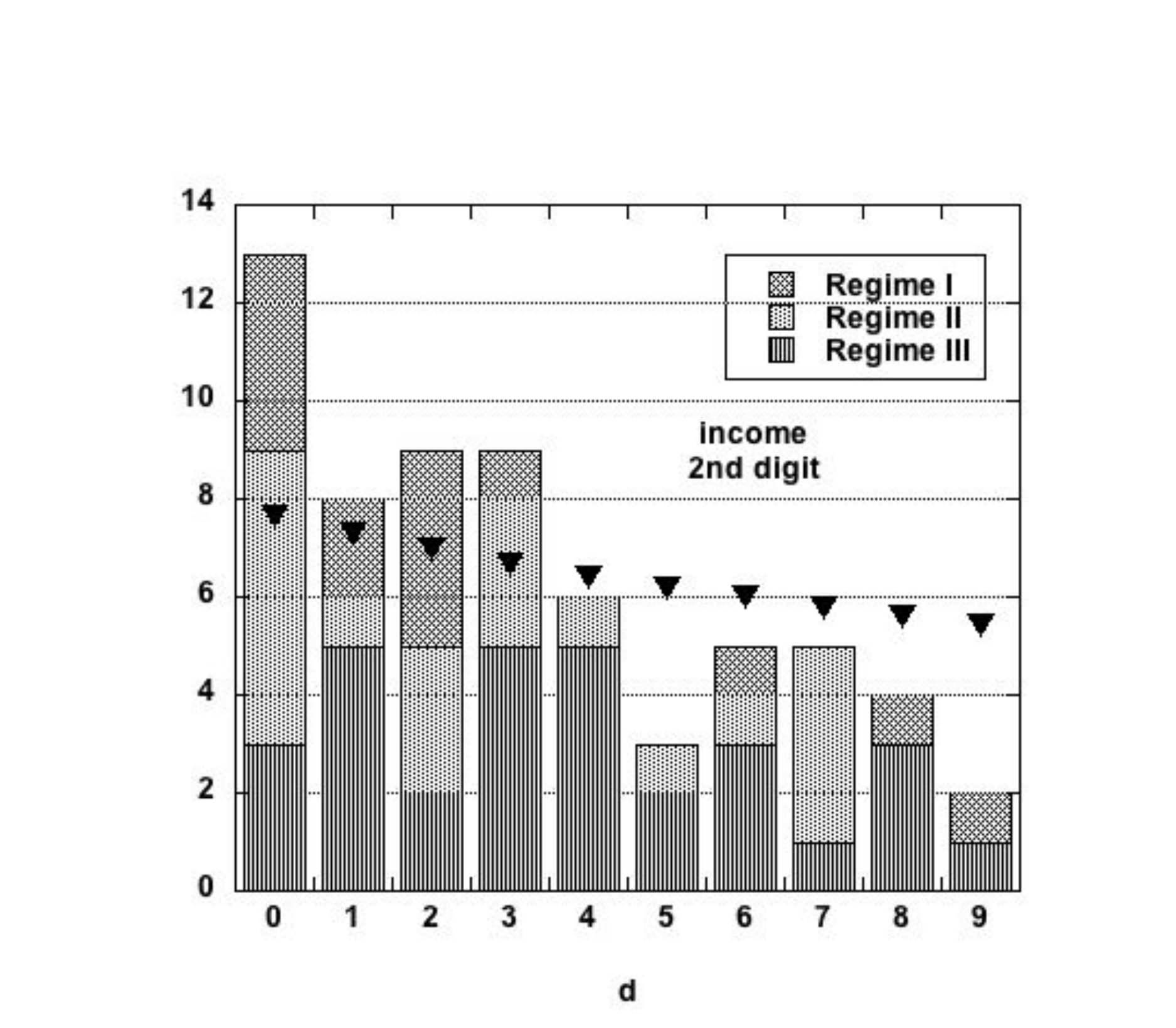}
\includegraphics[height=6.0cm,width=6.0cm]{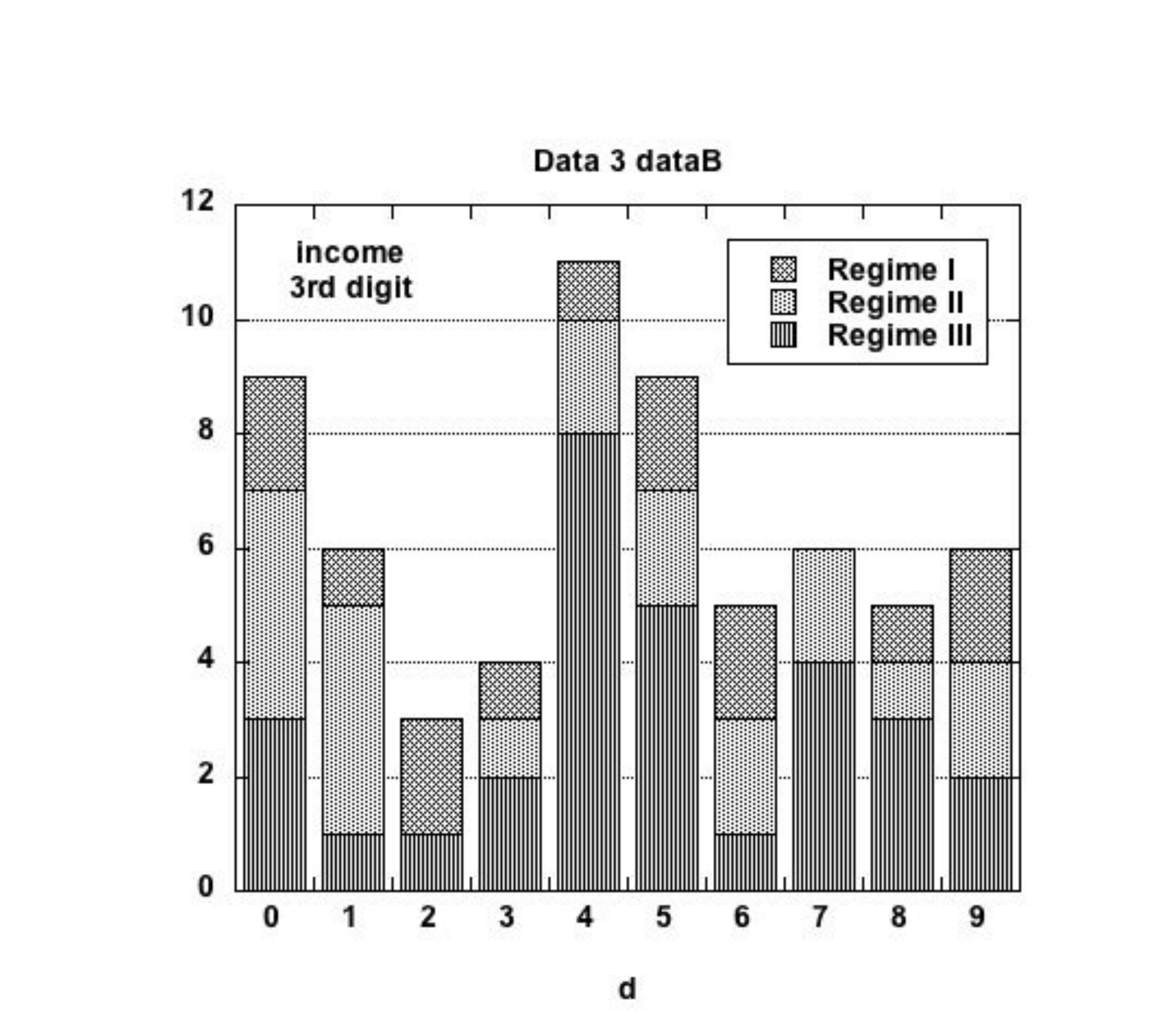}
\includegraphics[height=6.0cm,width=6.0cm]{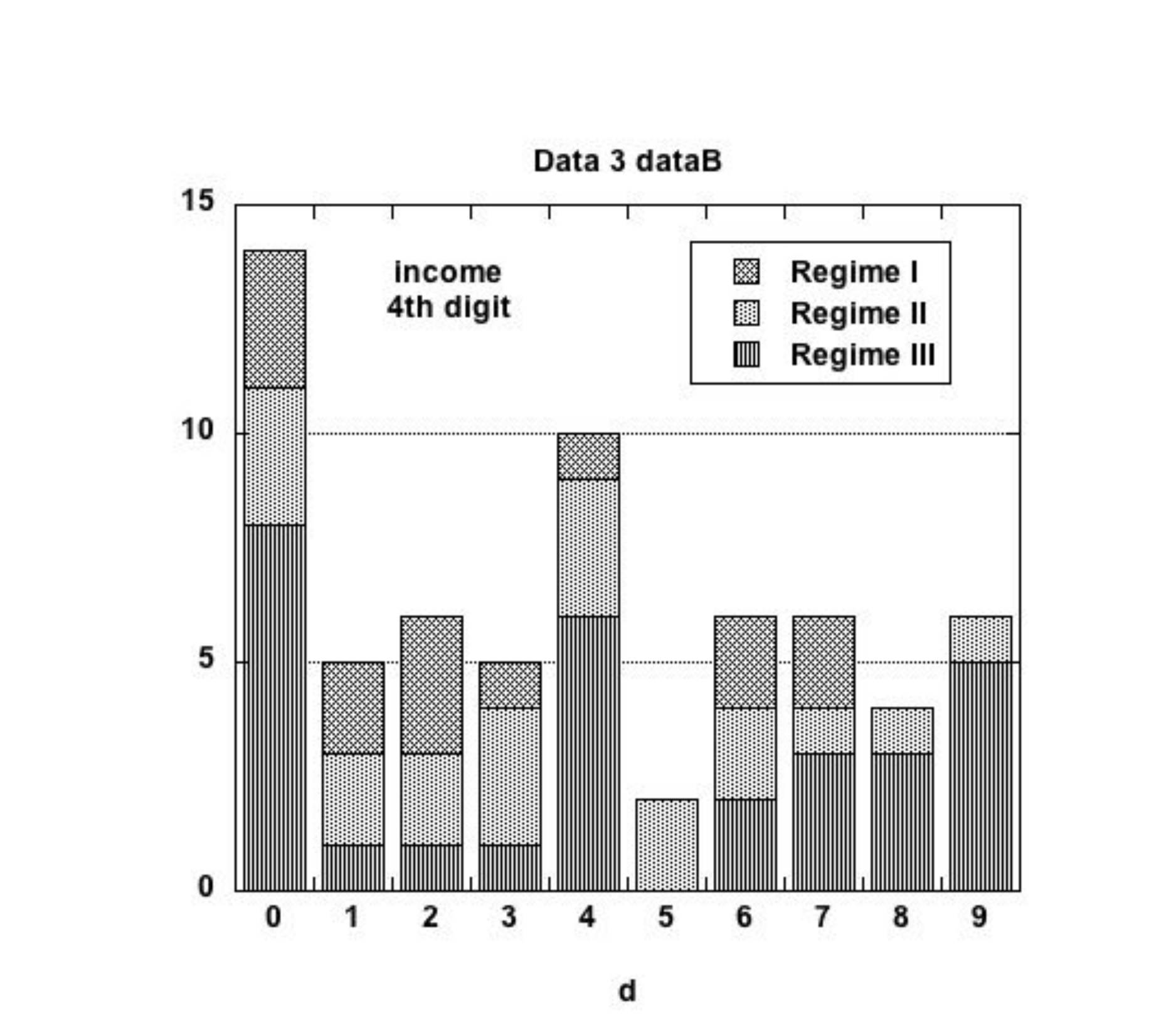}
\caption{    Benford histogram of the first four digits of  reported  income in the yearly budget of  the Belgian Antoinist community: the previously found three growth-decay time interval regimes are distinguished and summed up.  The expected Benford's laws for the first two digits are shown by  triangles}
\label{fig:Fig2inc}
\end{figure}

 \begin{figure} 
\centering
\includegraphics[height=6.0cm,width=6.0cm]{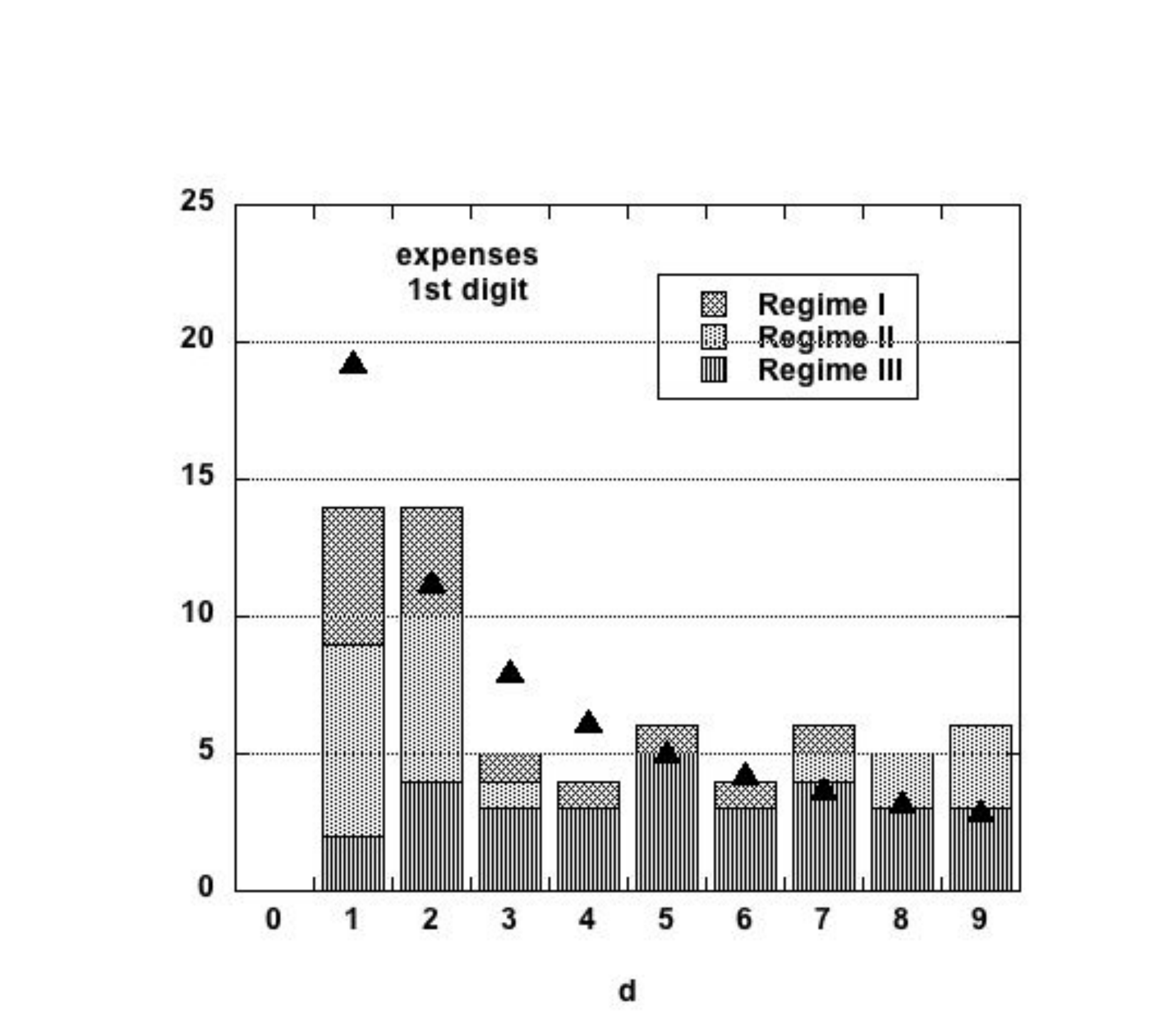}
\includegraphics[height=6.0cm,width=6.0cm]{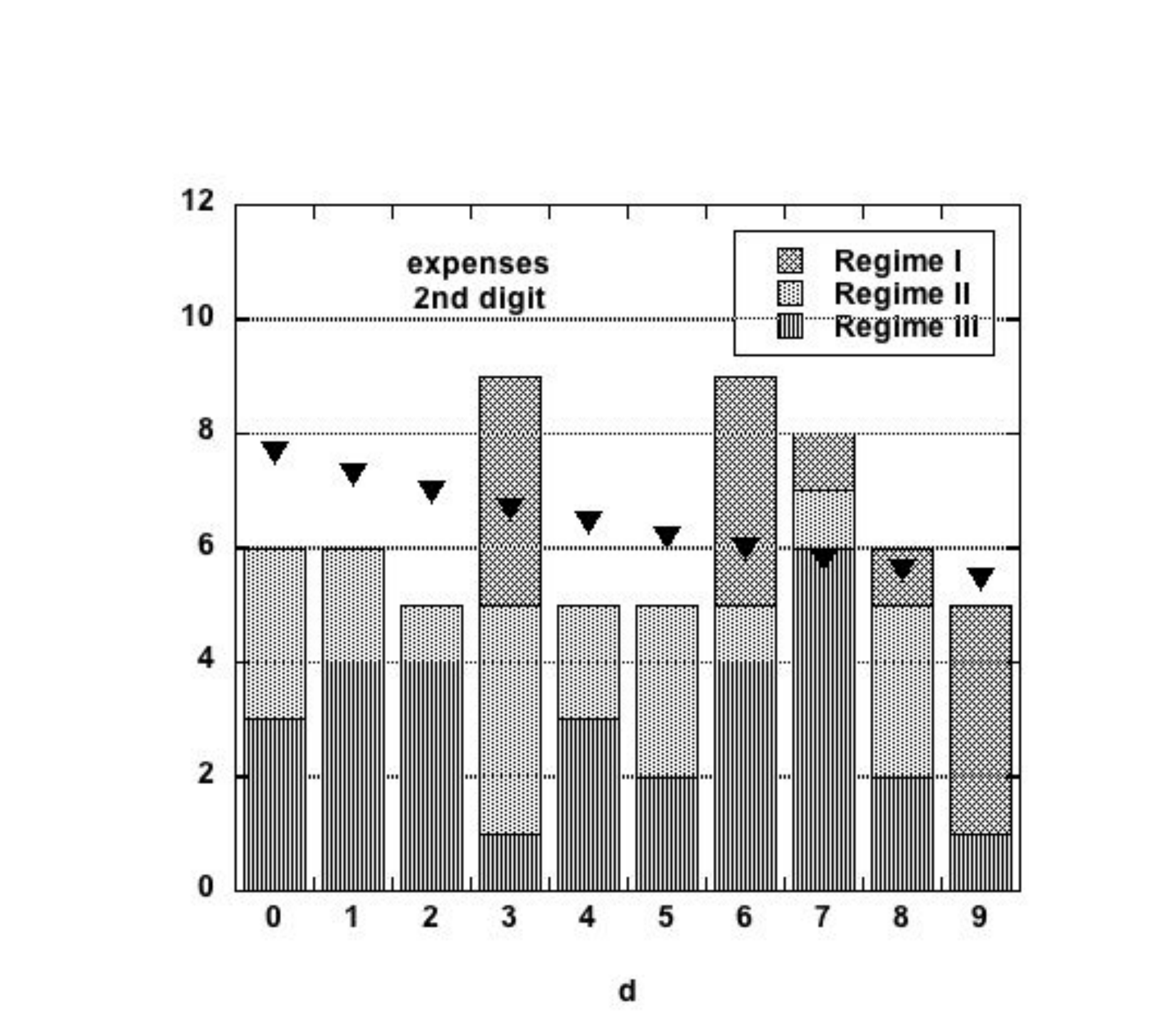}
\includegraphics[height=6.0cm,width=6.0cm]{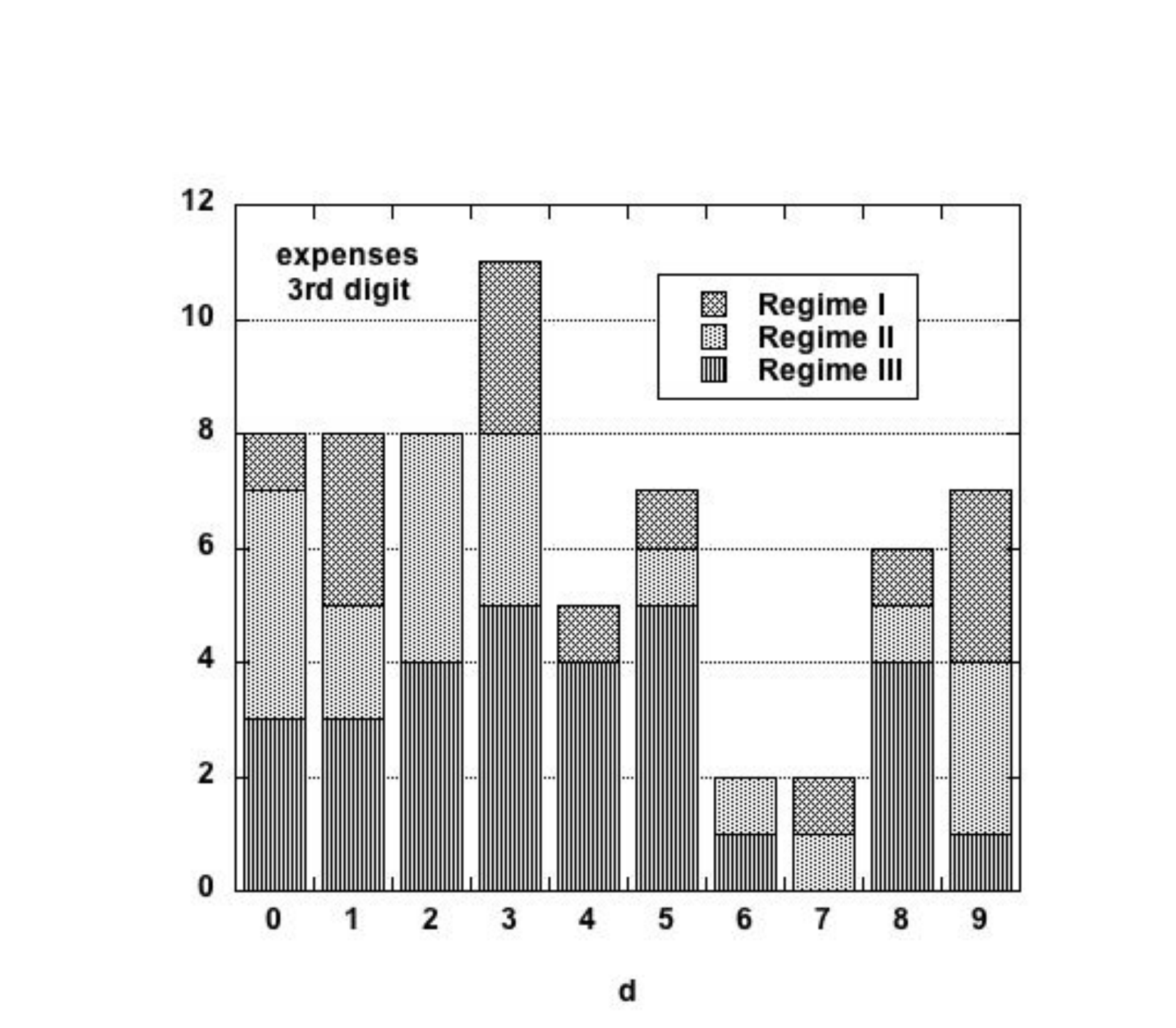}
\includegraphics[height=6.0cm,width=6.0cm]{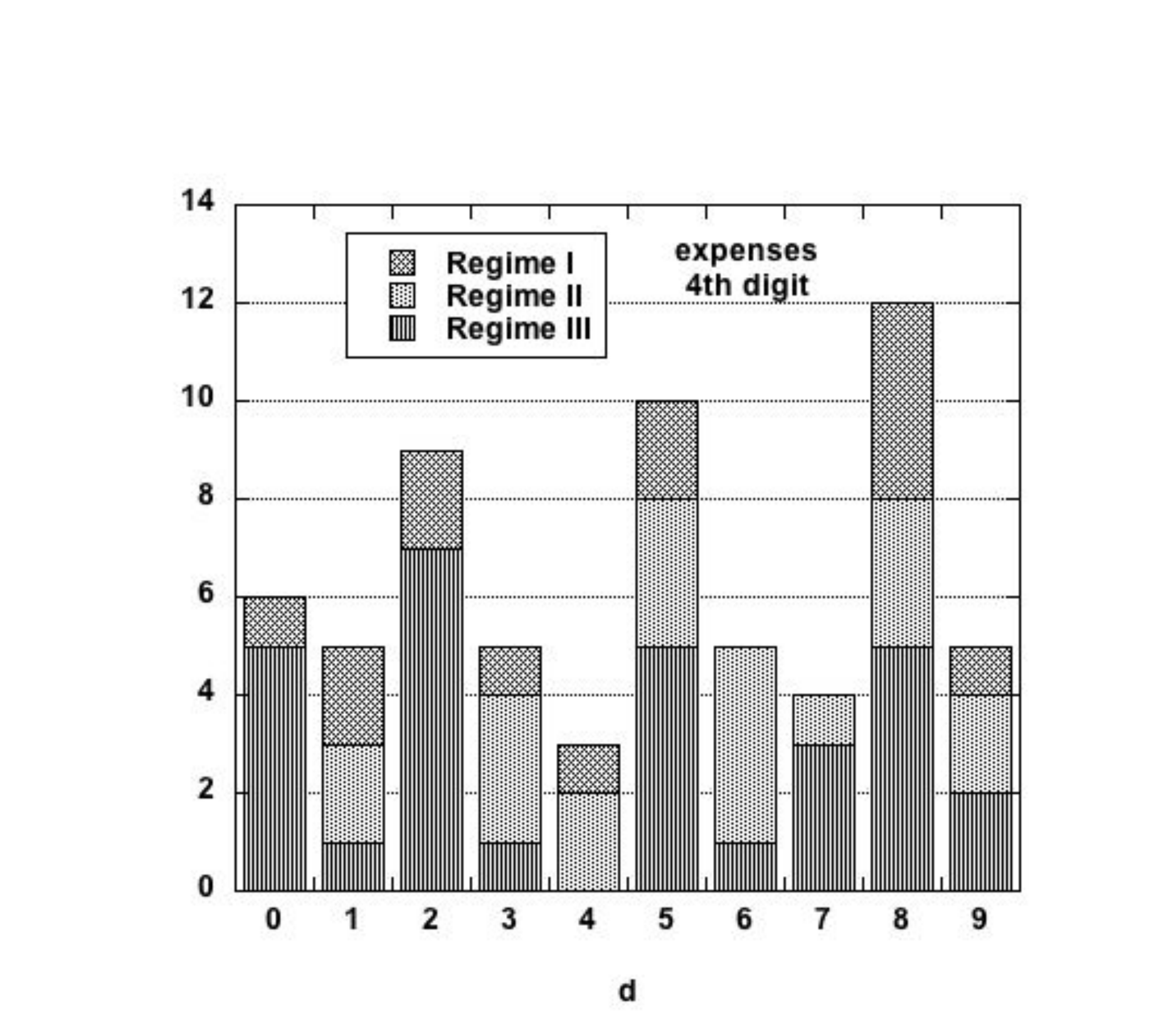}
\caption   {Benford histogram of the first four digits  of Theil  transformed   expenses in the yearly budget of  the Belgian Antoinist community: the previously found three growth-decay time interval regimes are distinguished and summed up.  The expected Benford's laws for the first two digits are shown by  triangles}
\label{fig:Fig3exp} 
\end{figure}

 \begin{figure}
\centering
\includegraphics[height=6.0cm,width=6.0cm]{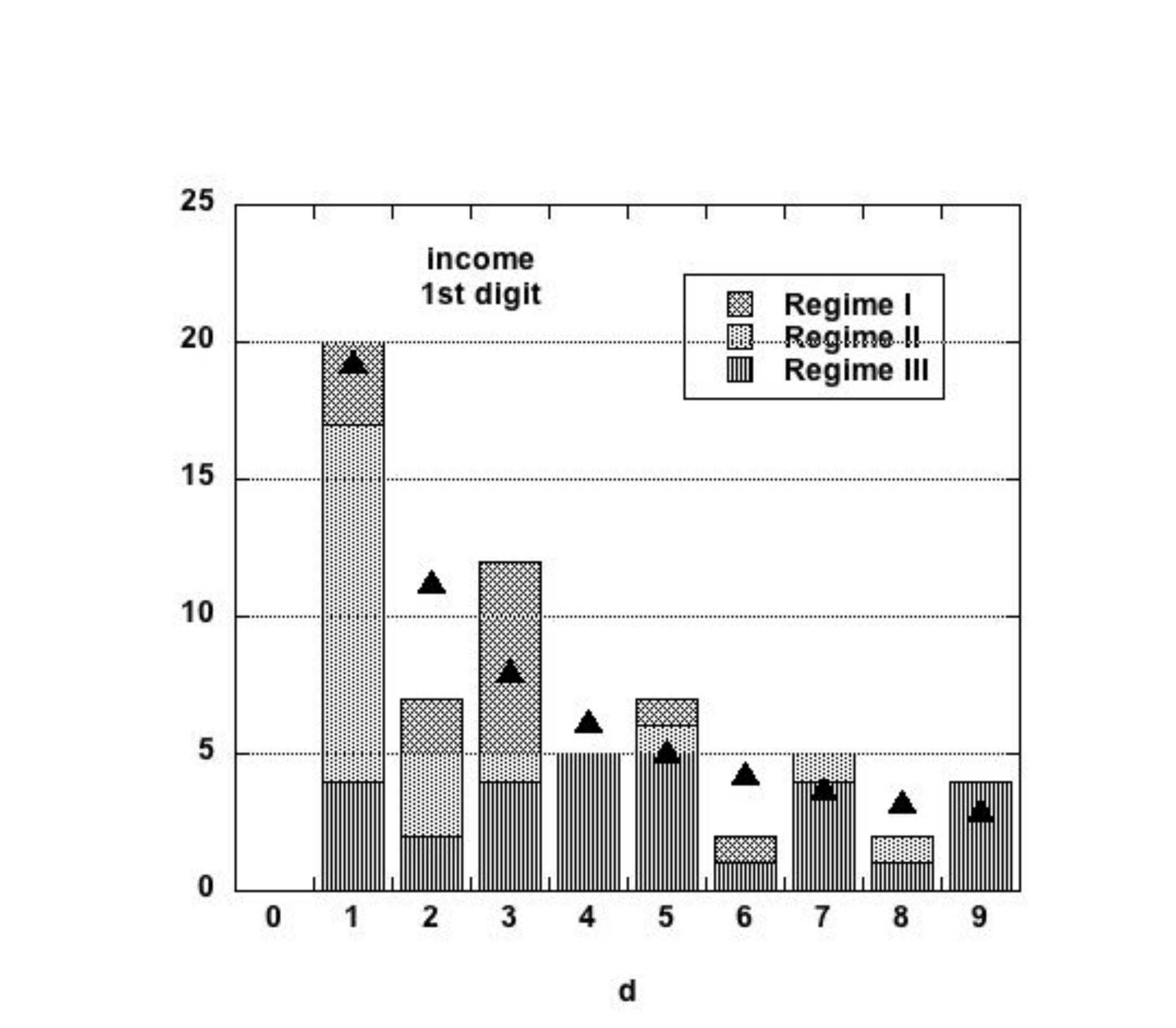}
\includegraphics[height=6.0cm,width=6.0cm]{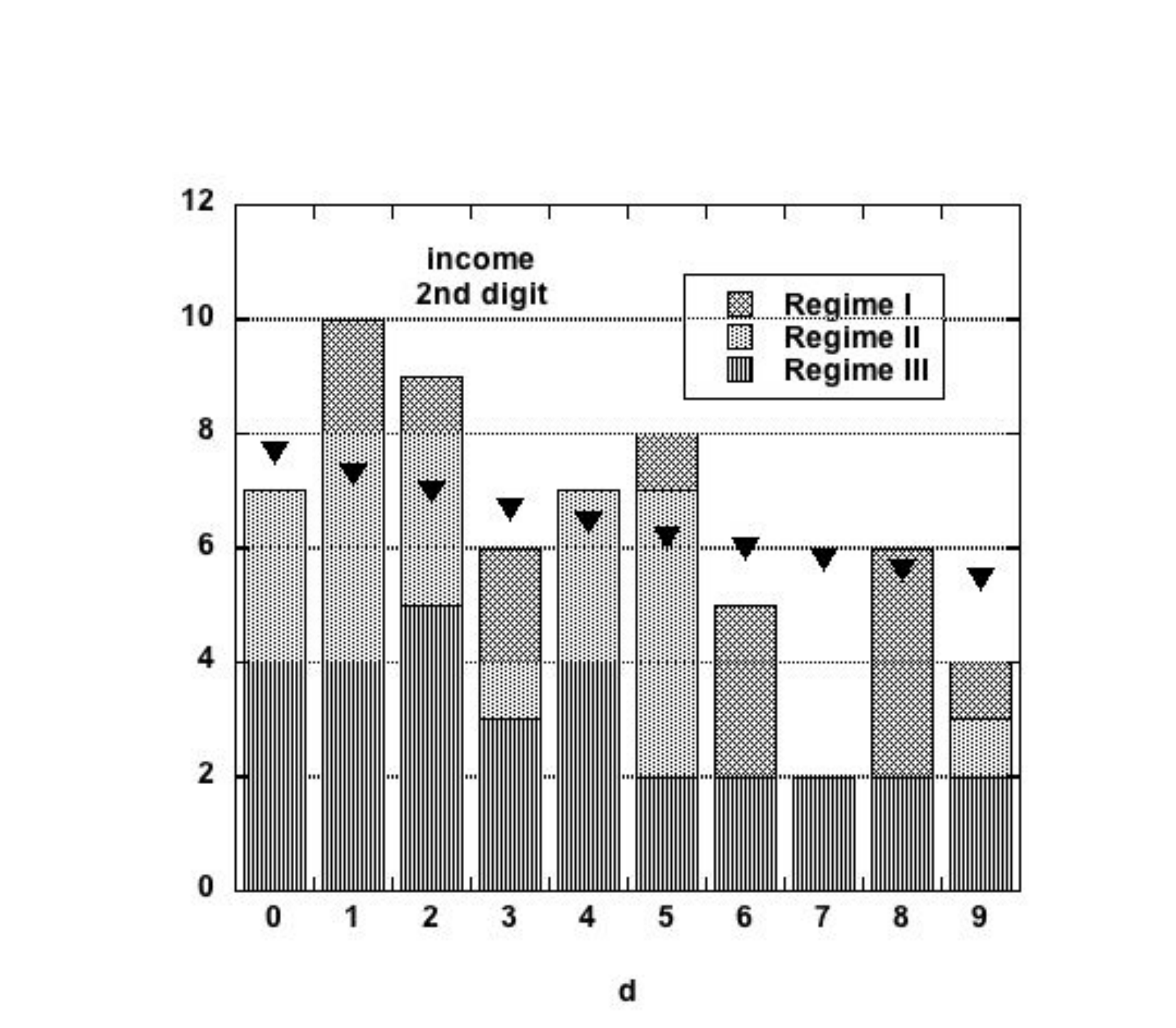}
\includegraphics[height=6.0cm,width=6.0cm]{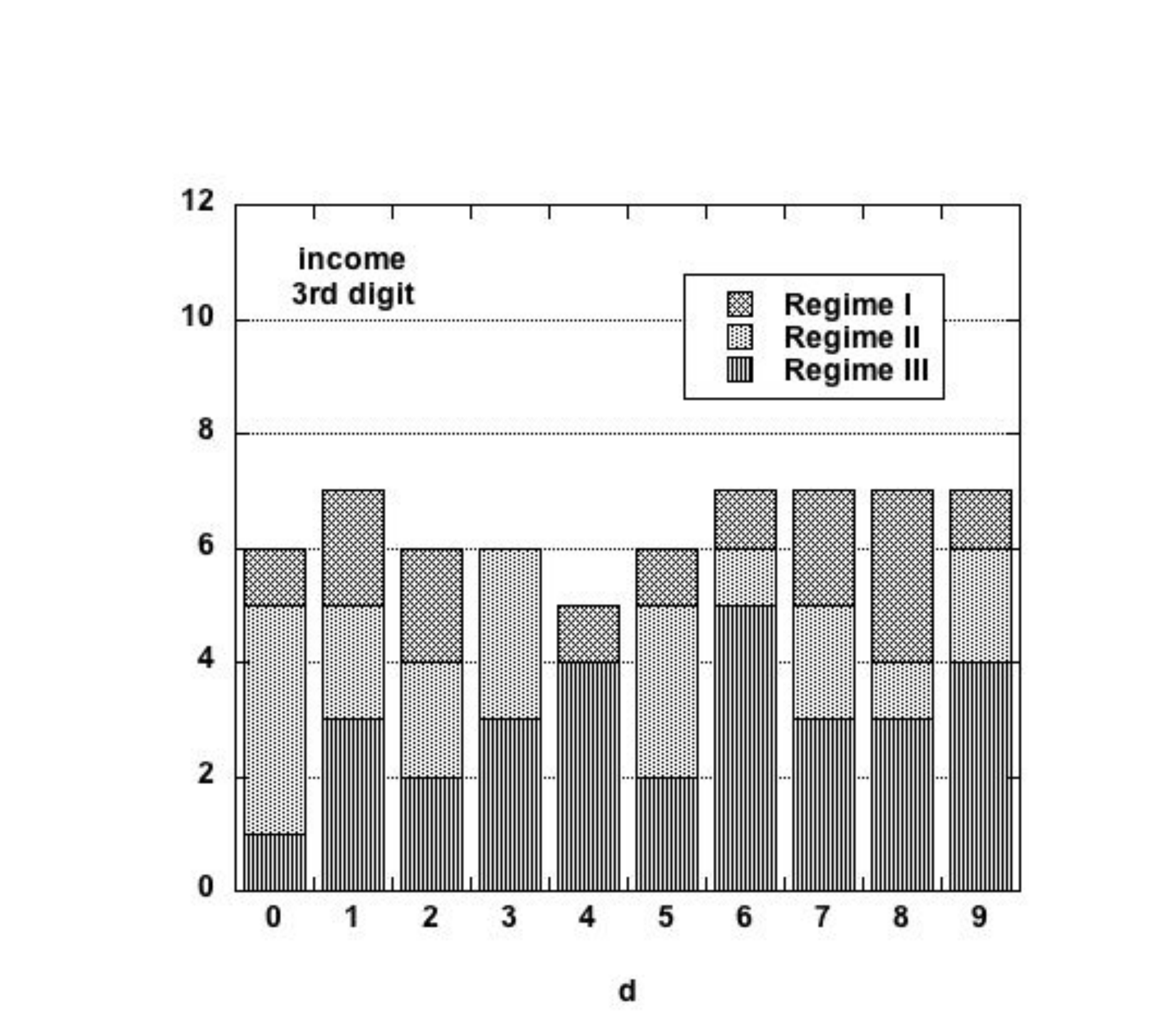}
\includegraphics[height=6.0cm,width=6.0cm]{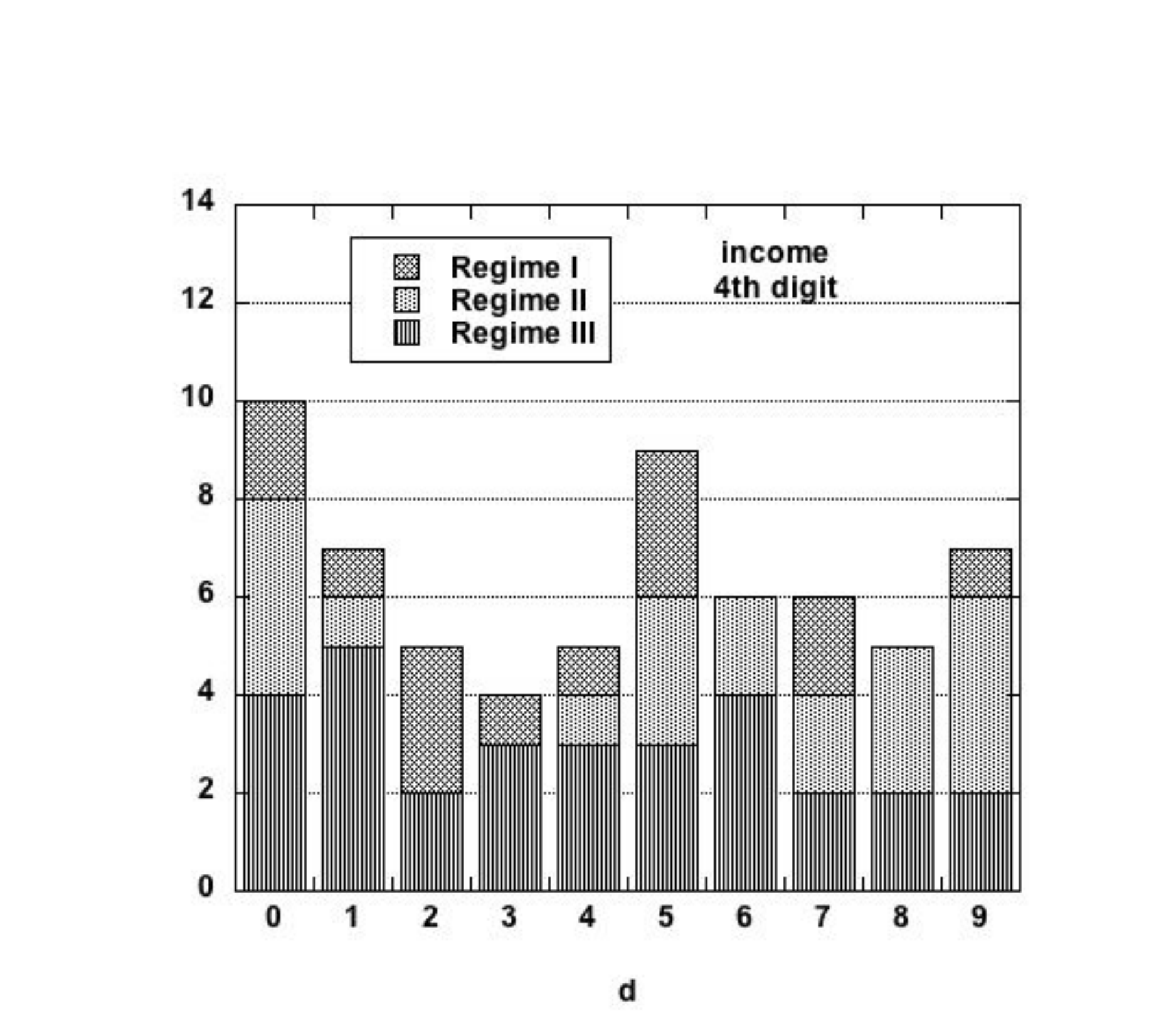}
\caption{    Benford histogram of the first four digits of  Theil  transformed  income in the yearly budget of  the Belgian Antoinist community: the previously found three growth-decay time interval regimes are distinguished and summed up.  The expected Benford's laws for the first two digits are shown by  triangles}
 \label{fig:Fig4inc}
\end{figure}

 \begin{figure}
\centering
\includegraphics[height=10cm,width=12cm]{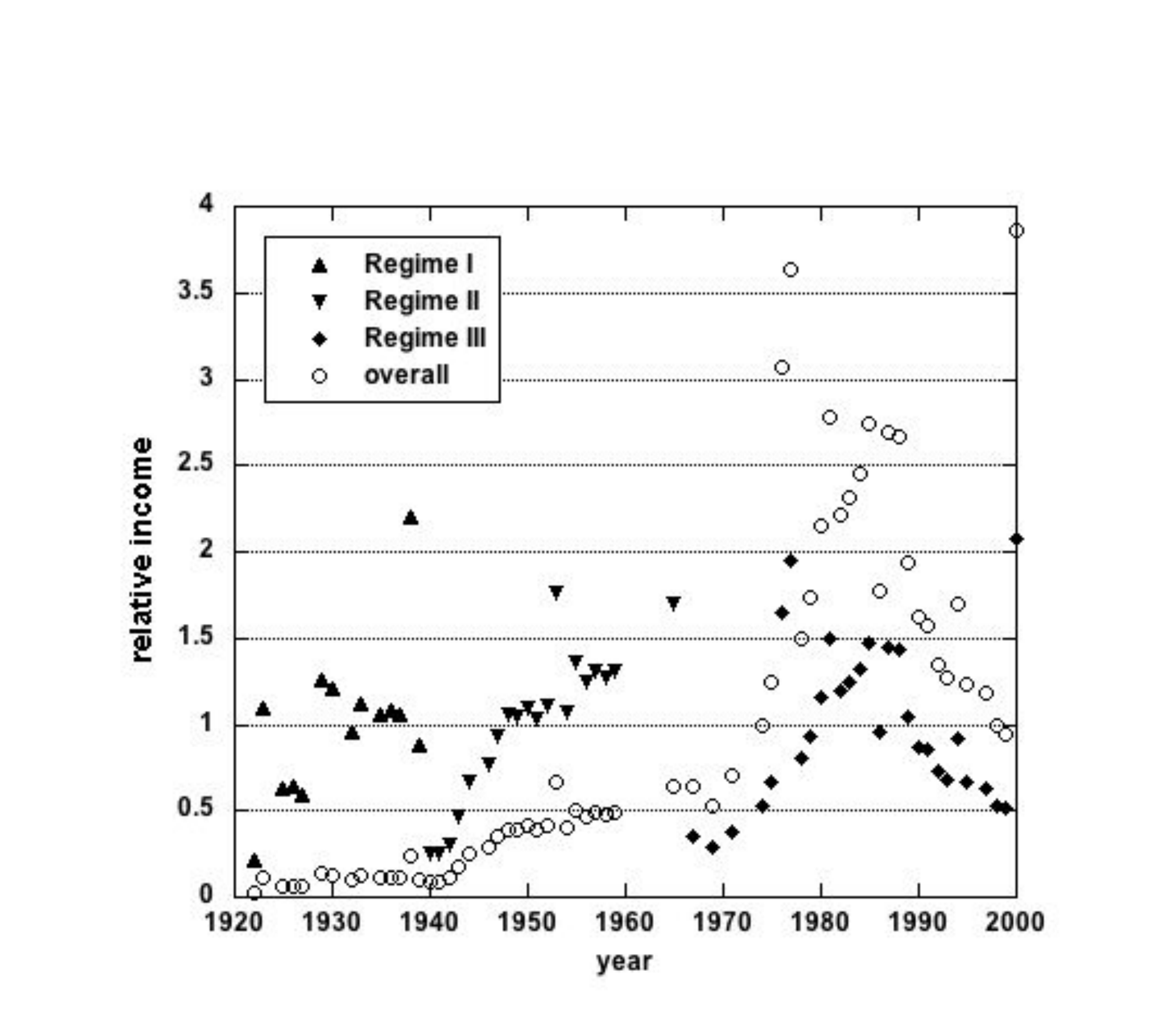}
\includegraphics[height=10cm,width=12cm]{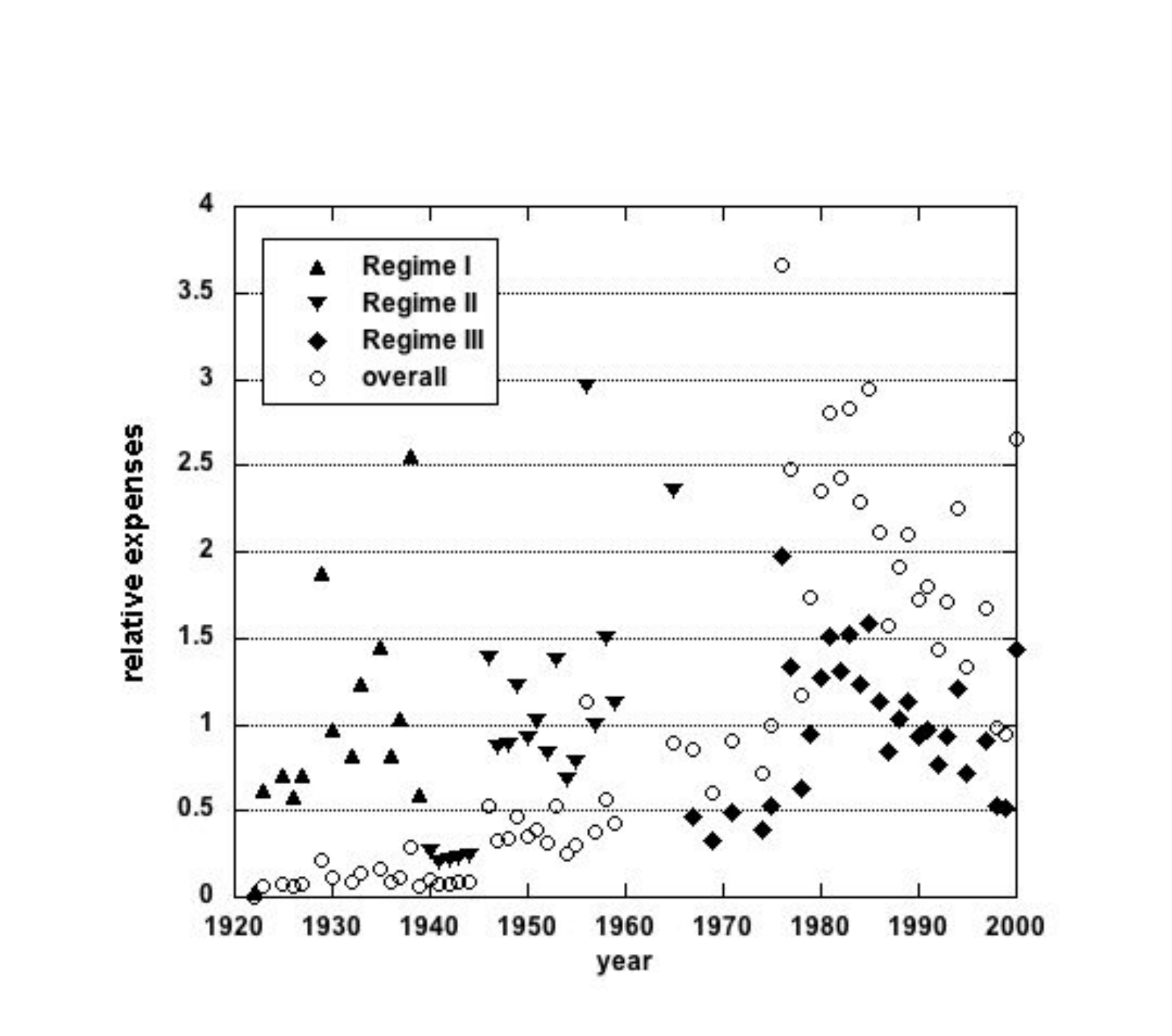}
\caption{Yearly relative income and expenses of the Belgium Antoinist community from the yearly budgets reported in the {\it Moniteur Belge}; three time regimes can be distinguished; the overall relative data is also shown }
\label{fig:relincomexpenses}
\end{figure}

\begin{figure}
\centering
\includegraphics[height=14cm,width=14cm]
{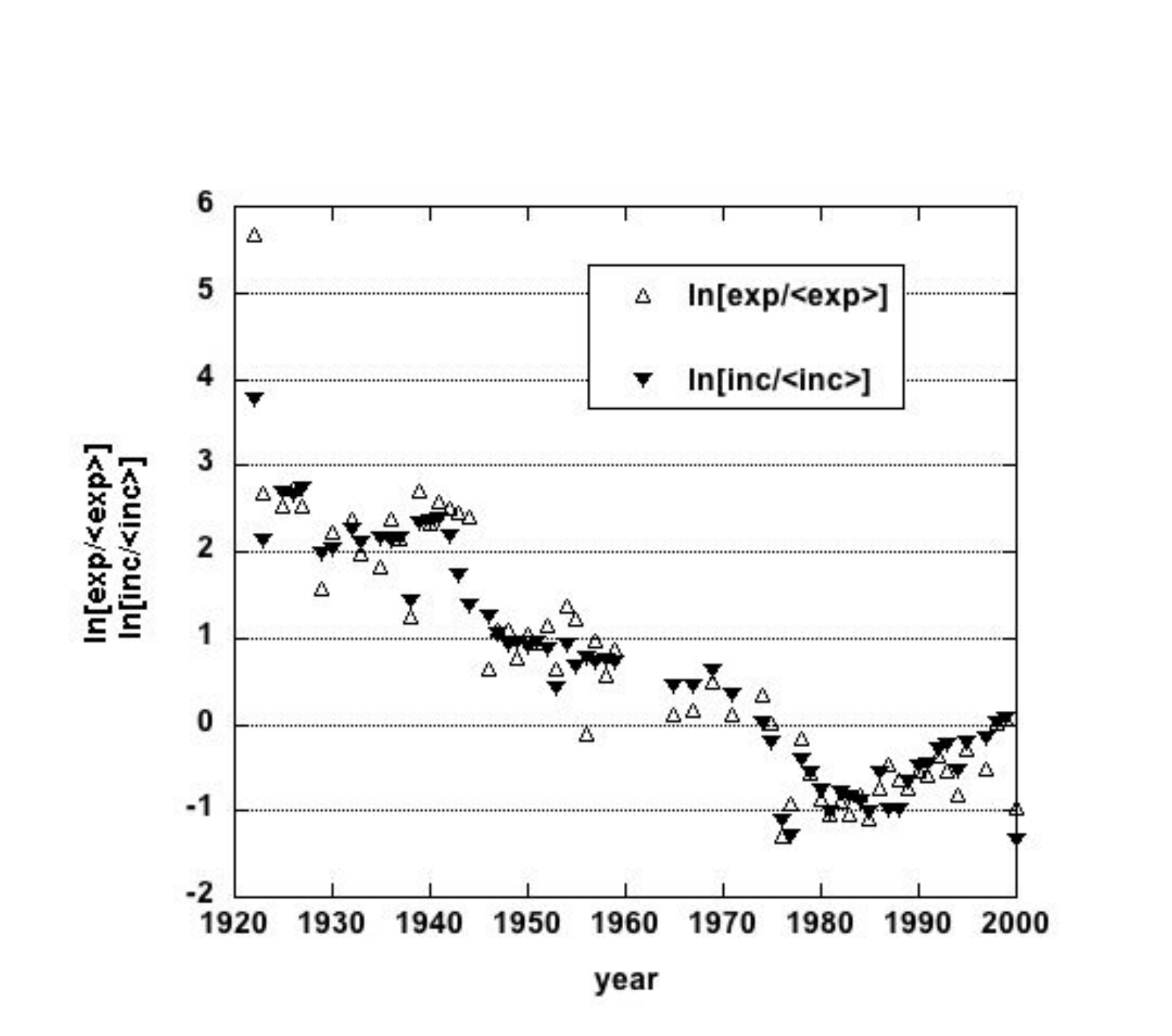}
\caption{Natural log of the yearly relative income (inc) and expenses  (exp) of the Belgium Antoinist community adapted from the yearly budgets reported in the {\it Moniteur Belge}; although the average expenses (<exp>) and income (<inc>) are taken over the whole time range, three time regimes can be distinguished  }
\label{fig:lnrelatincomlnrelatexpenses}
\end{figure}

\begin{figure}
\centering
\includegraphics[height=14cm,width=14cm]
{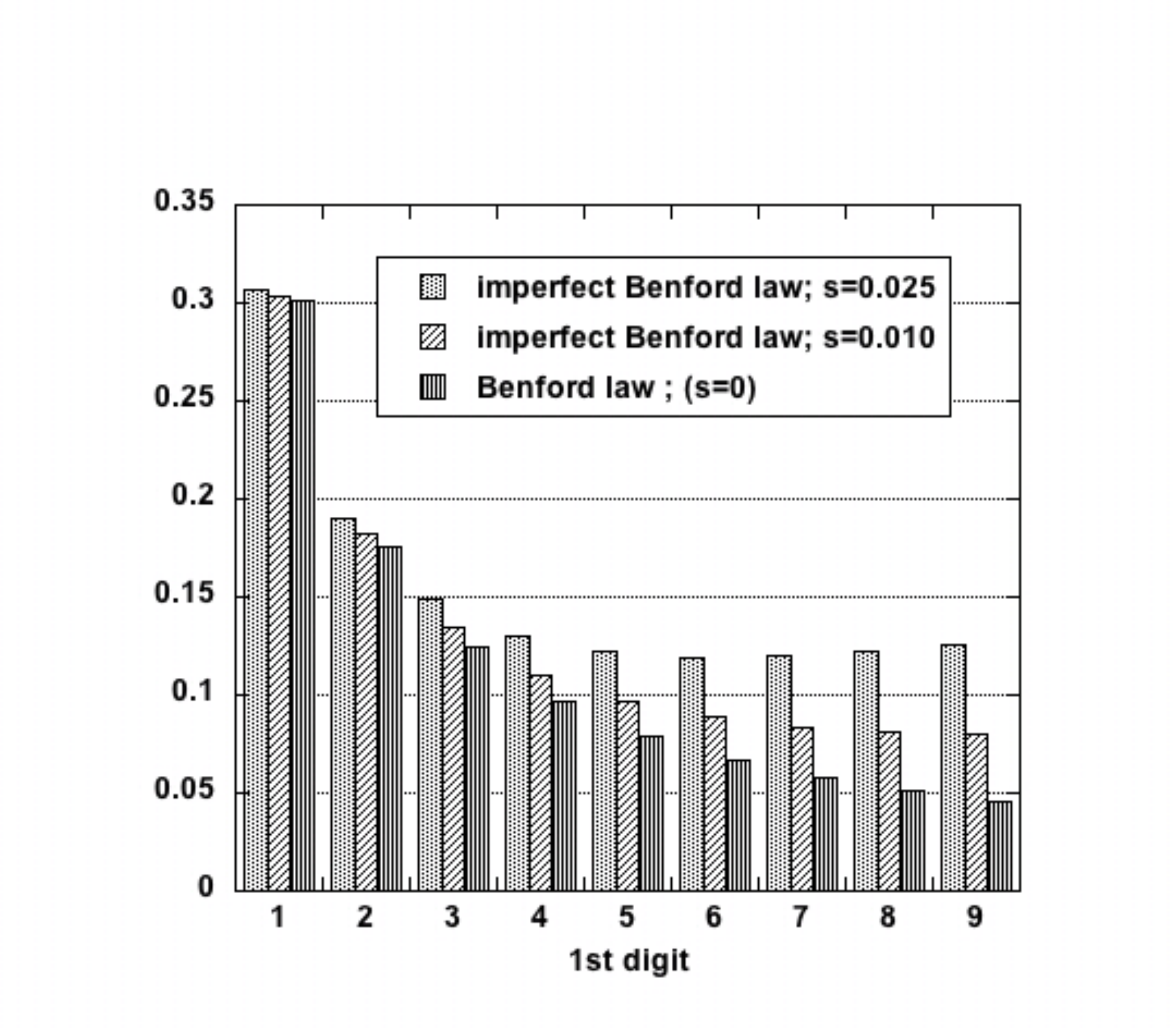}
\caption{Imperfect Benford law, Eq.(\ref{PCMA})  }
\label{fig:Plot7colBJIimperfBen1}
\end{figure}

 \begin{figure}
\centering
\includegraphics[height=6.0cm,width=6.0cm]{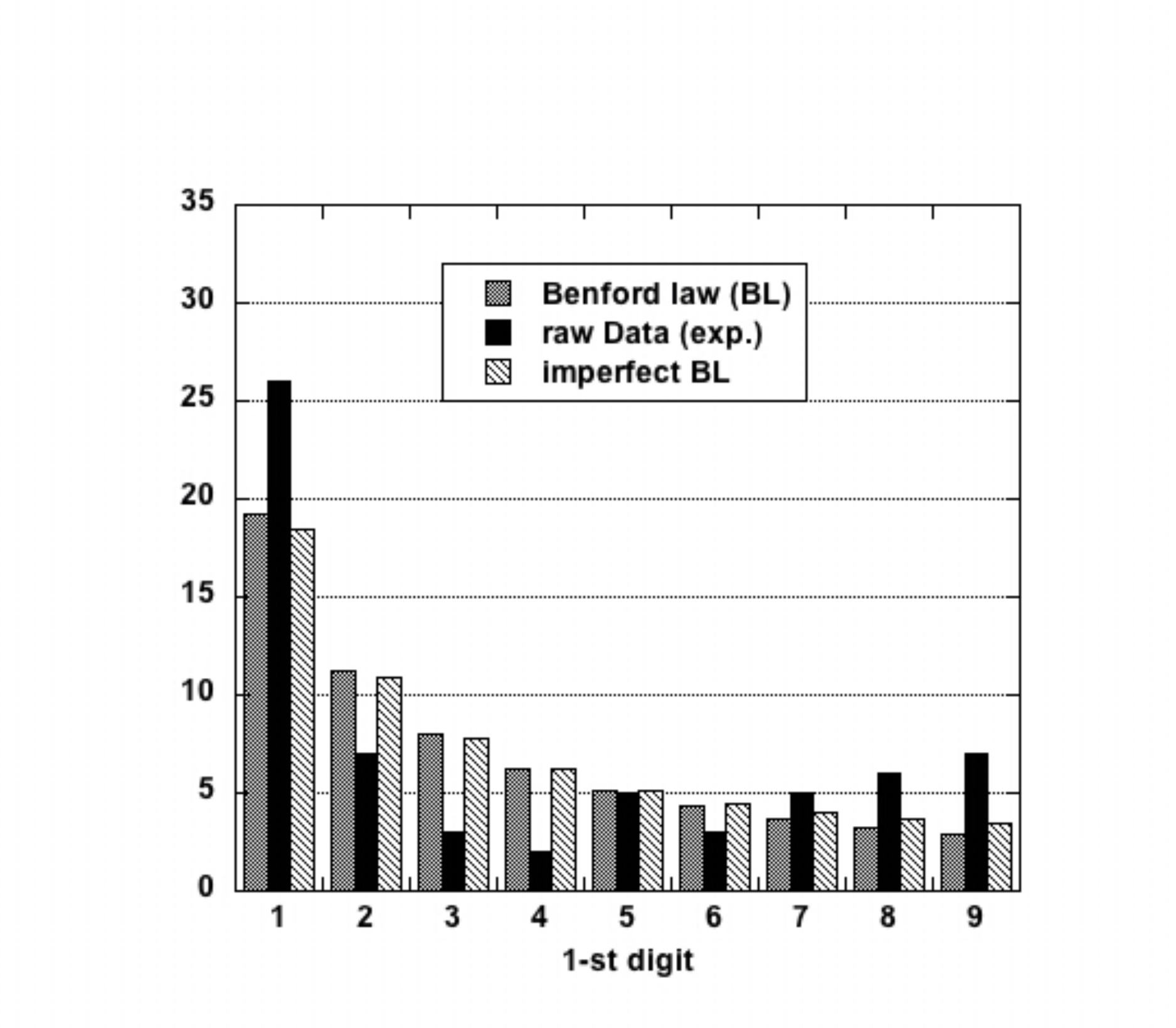}\includegraphics[height=6.0cm,width=6.0cm]{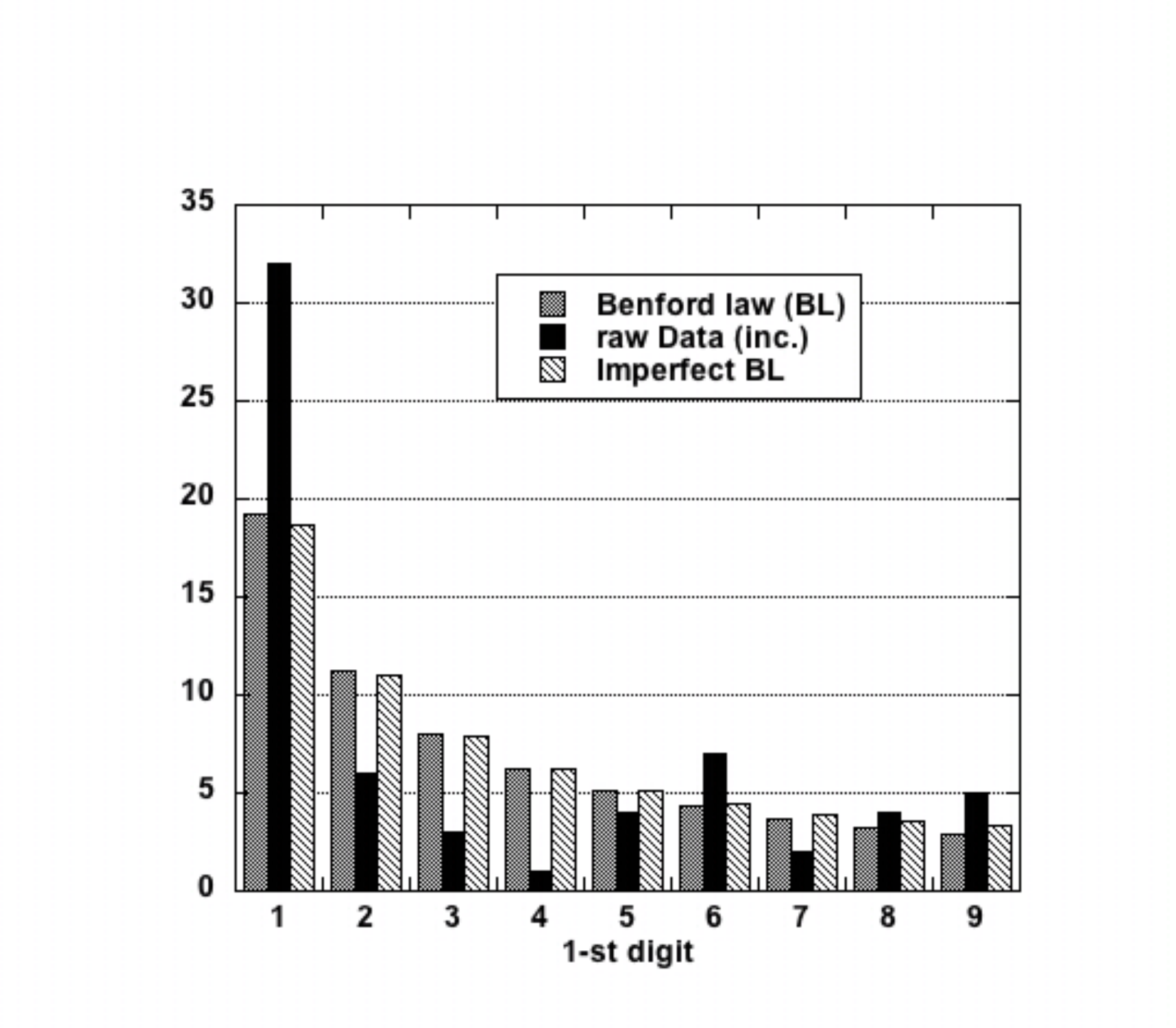}
\includegraphics[height=6.0cm,width=6.0cm]{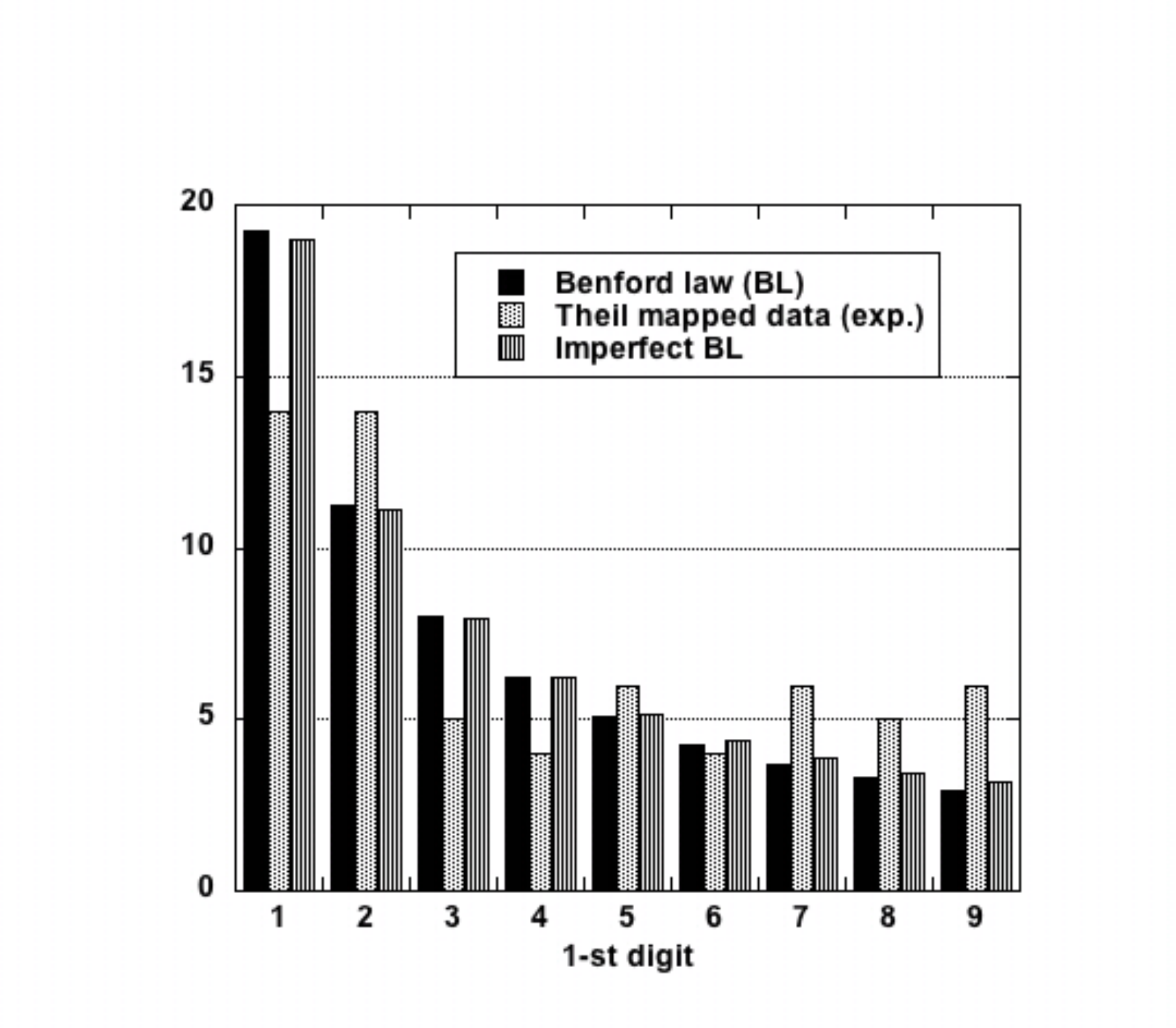}\includegraphics[height=6.0cm,width=6.0cm]{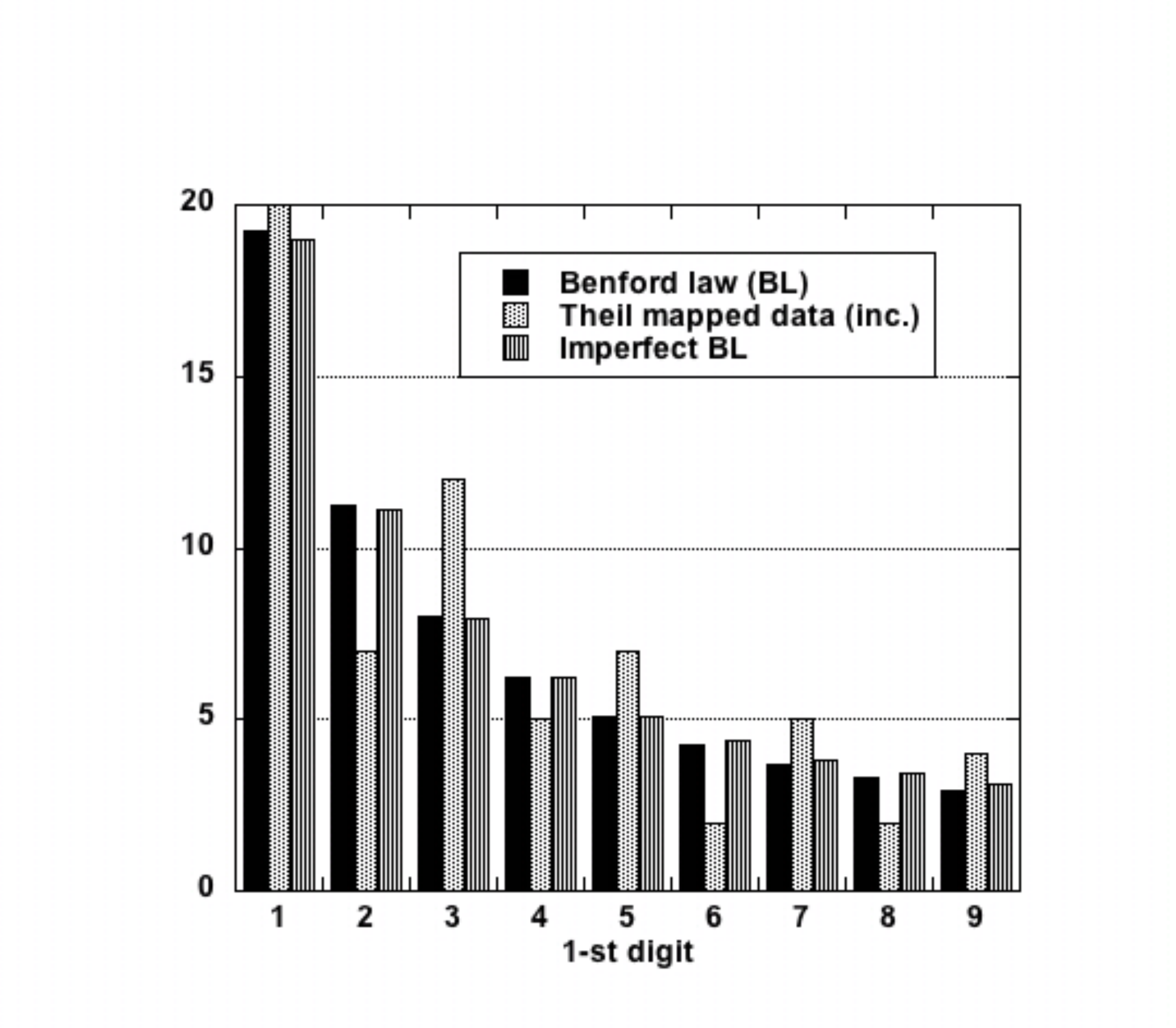}
\caption{    Plots of the  raw and Theil mapped data  histogram of the 1st-digit  in  the reported expenses  and   income for the yearly budget of  the Belgian Antoinist community comparing the expected  and   imperfect Benford's laws;  the best fit parameters of the latter are given in Table 3.}
 \label{fig:FigAppB}
\end{figure}

\end{document}